\documentclass[useAMS,usenatbib]{mn2e}
\usepackage{aasmacros}
\usepackage{times}
\usepackage{graphicx}
\usepackage{amssymb}
\usepackage[usenames,dvipsnames]{color}
\usepackage{url}
\usepackage{fixltx2e}
\usepackage{mathtools}
\usepackage{amsmath}
\usepackage{float}
\usepackage{cite}
\usepackage{fixltx2e}
\bibpunct{(}{)}{;}{a}{,}{,}
\usepackage{longtable}
\usepackage{hyperref}
\usepackage{morefloats}
\usepackage[T1]{fontenc}
\usepackage{ae,aecompl}

\let\cite=\citen

\def\apjs{ApJl}

\defcitealias{Condon12}{CO12}
\bibpunct{(}{)}{;}{a}{}{,}

\title[Deep 3-GHz Number Counts from a \textit{P(D)} Fluctuation Analysis]{Deep 3 GHz Number Counts from a \textit{P(D)} Fluctuation Analysis}

\author[Vernstrom et. al]{T. Vernstrom\thanks{E-mail:tvern@phas.ubc.ca}, Douglas Scott$^1$, J.V. Wall$^1$, J.J. Condon$^2$, W.D. Cotton$^2$,\newauthor
 E.B. Fomalont$^2$, K.I. Kellermann$^2$, N. Miller$^3$, R.A. Perley $^4$ \\
  $^1$Department of Physics and Astronomy, University of British Columbia, Vancouver, BC V6T 1Z1, Canada\\
  $^2$National Radio Astronomy Observatory, 520 Edgemont Rd, Charlottesville, VA 22903, USA\\
  $^3$Department of Astronomy, University of Maryland, College Park, MD, 20742, USA\\
  $^4$National Radio Astronomy Observatory, P.O. Box 0, Soccoro, NM, 87801, USA\\
}
\begin{document}
  

\pagerange{\pageref{firstpage}--\pageref{lastpage}} \pubyear{2013}

\maketitle

\label{firstpage}
\begin{abstract}

Radio source counts constrain galaxy populations and evolution, as well as the global star formation history. However, there is considerable disagreement among the published $1.4$-GHz source counts below $100\, \mu$Jy.  Here we present a statistical method for estimating the $\mu$Jy and even sub-$\mu$Jy source count using new deep wide-band $3$-GHz data in the Lockman Hole from the Karl G. Jansky Very Large Array (VLA). We analyzed the confusion amplitude distribution \textit{P(D)}, which provides a fresh approach in the form of a more robust model, with a comprehensive error analysis. We tested this method on a large-scale simulation, incorporating clustering and finite source sizes. We discuss in detail our statistical methods for fitting using Monte Carlo Markov chains, handling correlations, and systematic errors from the use of wide-band radio interferometric data. We demonstrated that the source count can be constrained down to $50\,$nJy, a factor of 20 below the rms confusion. We found the differential source count near $10\, \mu$Jy to have a slope of $-1.7$, decreasing to about $-1.4$ at fainter flux densities. At $3\,$GHz the rms confusion in an $8\,$arcsec FWHM beam is $\sim 1.2\, \mu$Jy beam$^{-1}$, and a radio background temperature $\sim 14\,$mK. Our counts are broadly consistent with published evolutionary models. With these results we were also able to constrain the peak of the Euclidean normalized differential source count of any possible new radio populations that would contribute to the cosmic radio background down to $50\,$nJy.

\end{abstract}


\section{Introduction}
\label{sec:introduction}
The counts of discrete radio sources and their contribution to the cosmic radio background (CRB) can be used to constrain galaxy evolution. The history of radio source counts goes back to \citet{Mills52} who discussed the cumulative count or `ogive', giving one of the first published source counts, followed by \citet{Ryle55a}, with the discovery that the slope of the source count using the confusion \textit{P(D)} distribution was steeper than expected from a static Euclidean universe, which implied that sources must be evolving in space density or luminosity. Since then many more surveys have been carried out to measure the source counts at various radio frequencies, both whole-sky and limited area surveys, which confirmed the importance of evolution \citep[see][for a recent review]{dezotti09}. These counts can now be broken down into different source populations, primarily star forming galaxies and those powered by active galactic nuclei. 

With advancements in radio telescopes we have been able to probe the source count to ever increasing depth, so that estimating the count in the $\mu$Jy and sub-$\mu$Jy regions is now possible. Investigating the count at these faint flux densities is important for understanding the evolution of sources at lower luminosities and/or higher redshifts ($z \geq 2$). How the count below $10 \, \mu$Jy behave has been unknown until now: what the slope is in this region, whether the Euclidean-normalised count declines or begin to rise again (possibly indicating a new population), and whether they continue to obey the well-known far-IR to radio correlation \citep{deJong85,Condon91,Ivison10} out to higher redshifts. 

There has been considerable discussion recently about what might be happening at flux densities fainter than the limits of the current source counts. Results from the Absolute Radiometer for Cosmology, Astrophysics, and Diffuse Emission, or ARCADE 2 \citep{Fixsen09}, revived interest in the CRB and how it may be related to the faint\footnote{In this paper the terms``faint" and ``bright" refer to the flux density of the source or sources.}  counts. Measurements from ARCADE 2 indicate the presence of an excess of radio emission over previous measurements or estimates using source count data \citep{Seiffert09}. \citet{Vernstrom11}, motivated by the ARCADE 2 results, presented new estimates of lower limits to the background from a compilation of source counts at eight frequencies and found an expected value for the background temperature almost five times lower than that of ARCADE 2 at $1.4\,$GHz.  
The ARCADE2 results \citep{Seiffert09} suggested that this excess emission might be coming from a previously unrecognized population of discrete radio sources below the flux density limit of existing surveys, and that this new population might be seen in radio source counts extending to lower flux density levels. This issue was further examined by \citet{Singal10}, who concluded that this emission could primarily be coming from ordinary star forming galaxies at $z >$ 1 \textit{only if} the far-IR/radio ratio decreases with redshift. In other words we can only explain the background results with sources if they break the far--IR/radio correlation \citep{Haarsma98}. 

\citet{Vernstrom11} also showed that the known radio source counts cannot on their own account for the ARCADE 2 excess, although the source counts, at least at $1.4\,$GHz, are not inconsistent with a possible upturn below about $10 \, \mu$Jy. Such a possible upturn is mainly driven by the faintest count available at $1.4 \,$GHz, from \citet{Owen08}. Owen $\&$ Morrison found that their (Euclidean-normalised) count, which extends down to $15 \, \mu$Jy, did not decrease with decreasing flux density (compared to static Euclidean counts), but seemed to level off or even show signs of increasing. It is important to note that the Euclidean-normalised count ($S^{5/2}dN/dS$) does not need to level off or turn up to explain the high ARCADE 2 background temperature; it is sufficient that $S^2dN/dS$ levels off or turns up.

New $3$-GHz data from the Karl G. Jansky Very Large Array, VLA, reach down to $\mu$Jy levels \citep[][hereafter CO12]{Condon12} and the resulting map is the deepest currently available. In this previous paper we estimated the source count from $1$ to $10 \, \mu$Jy using a technique known for historical reasons as \textit{P(D)} analysis \citep{Scheuer57,Condon74}. This approach allows a statistical estimate of the count from a confusion-limited survey, extending down to flux densities below the confusion limit. \textit {P(D)} is the probability distribution of peak flux densities in an image. This approach results in statistical estimates of the source count that are much fainter than the faintest sources that can be counted individually (about 5 times the rms noise). The count model used in \citetalias{Condon12} was a single power law over a limited flux density range. However, there appeared to be evidence for a break in the slope somewhere in this region and certainly the results did not support any upturn in the count. While this previous result puts strong limits on the $\mu$Jy count, it is possible that more comprehensive analysis of the \textit{P(D)} distribution, with a more general count model, could reveal additional information about the true shape of the count, as well as constraining the count fainter than $1 \, \mu$Jy. 

Here we present a more sophisticated modelling approach to the \textit{P(D)} fitting process, motivated by \citet{Patanchon09}, using a model based on multiple joined power laws. The statistical uncertainties here are evaluated using Markov chains. We test this technique with a large-scale simulation incorporating realistic source sizes, multi-component sources, and clustering. This method allows for exploration of the flux density limit of the \textit{P(D)} approach, and the count below the confusion noise, as well as a thorough non-parametric error analysis. 

In Section~\ref{sec:obs} we briefly describe the data used. In Section~\ref{sec:method} we describe the details of \textit{P(D)} and the process adopted for model fitting and error analysis. In Section~\ref{sec:model} we discuss the models used in the fitting and the details of their application to the VLA data. Section~\ref{sec:sim} presents a discussion of the simulation used to test the method and the results from fitting the simulated data. In Section~\ref{sec:results} we present the results of the fitting from two different models, a discussion of the parameter degeneracies, and the derived radio background temperature. Section~\ref{sec:dis} gives a discussion of the systematics and comparisons with previous results.

\section{Observations}
\label{sec:obs}
The observations were made with the Karl G. Jansky Very Large Array (VLA) in $S$-band, which ranges from 2 to $4\,$GHz, in the C configuration $($maximum baseline 3.4 km$)$, with an average of 21 antennas. The $3$-GHz VLA pointing was selected explicitly to overlap the region Owen $\&$ Morrison (2008) observed in the Lockman Hole at $1.4\,$GHz. The field is centred on $\alpha = 10^{\rm{h}}46^{\rm{m}}00^{\rm{s}}$, $\delta = +59^{\circ}01'00''$ (J2000), and was originally chosen as it is known to be a ``random" (i.e. for our purposes quite crowded) field, with the brightest source about $7\,$mJy, and no very bright radio sources nearby. It is also covered in many other wavebands ({\it Spitzer, Chandra, Herschel}, GMRT, and more) allowing for source cross-identification, investigation of AGN contribution, and study of the far-IR/radio correlation. The $3\,$GHz (centre frequency) S-band was chosen rather than the $1.4\,$GHz L-Band, because the contamination from interference is less, the sensitivity is better, the requirements on dynamic range are lower, the confusion is lower, and additionally S-Band has greater available bandwidth ($2 \, $GHz). This is in addition to S-band being closer to the frequency of the largest ARCADE 2 observed excess.  

There was a total of approximately $50$ hours of on-source observing time, split into six $10$-hour sessions in 2012 February and March. The calibration and editing were performed using the \textsc{Obit} package \citep{Cotton08}\footnote{\url{http://www.cv.nrao.edu/~bcotton/Obit.html}}, and are described in detail in \citetalias{Condon12}. The VLA S-band contains 16 separate frequency sub-bands, which were cleaned simultaneously and then used to create a $3$-GHz wide-band image. The sub-bands were weighted to maximize signal-to-noise for sources having spectral indices near $\alpha = -0.7$, the mean spectral index of faint sources found at frequencies around $3 \,$GHz \citep{Condon84b}. The final wide-band image, and the 16 sub-band images, have circular $8\,$-arcsec synthesized beams. The FWHM of the primary beam ranges from $21.6\,$arcmin at $2\,$GHz to $ 10.8\,$arcmin at $4\,$GHz. Because of the weighting used to combine the individual images, the primary beam of the wide-band image is frequency-dependent, so the effective frequency $\langle\nu\rangle$ of the image decreases with radial distance from the pointing centre. Table~\ref{tab:image} provides a summary of the image properties. 

\begin{table}
\caption{Image properties for the wide-band VLA data. The reported noise values are all after correction for the primary beam and frequency weighting effects, with $\rho$ being the distance from the pointing centre. The clean beam size, $\theta_{\rm b}$, is the Full Width Half Max, FWHM, and the synthesized beam solid angle, $\Omega_{\rm b}$, is ($\theta_{\rm b}^2\pi)/(4{\rm ln}2$).   }
\begin{tabular}{lll}
\hline
\hline
Quantity & Value & Unit \\
\hline
$\langle\nu\rangle$ in centre & 3.06 & GHz\\
$\langle\nu\rangle$ at 5 arcmin & 2.96 & GHz\\
$\langle\nu\rangle$ inside 5 arcmin Ring &3.02 & GHz\\
Pixel size & 1.252 & arcsec\\
Clean beam FWHM, $\theta_{\rm b}$ & 8.00 & arcsec\\
Beam solid angle, $\Omega_{\rm b}$ & 72.32 &arcsec$^2$ \\
$\sigma_{\rm n}(\rho$=0) & 1.08 & $\mu$Jy beam$^{-1}$ \\
$\sigma_{\rm{n}}(\rho=5\prime$)& 1.447 & $\mu$Jy beam$^{-1}$ \\
$\sigma_{\rm{n}}(\rho \le5\prime$)& 1.255 & $\mu$Jy beam$^{-1}$ \\
\hline
\end{tabular}
\label{tab:image}
\end{table}

\section{Method}
\label{sec:method}

In order to model the source count below the current cut-offs, the method of \textit{P(D)}, or probability of deflection, is used. This method was introduced by \citet{Scheuer57} as the probability of pen deflections on a chart-recorder from a single baseline of a two-element radio interferometer. The \textit{P(D)} distribution of an image is the distribution of pixel intensities (Jy beam$^{-1}$), or the ``1-point statistics", which depends on the underlying source count. \citet{Condon74} and \citet{Scheuer74} gave analytical derivations of \textit{P(D)} for a single power-law model of a source count. The method which has been most often applied is to count the objects in the map brighter than some cut-off (usually about 5$\sigma_{\rm n}$) and use \textit{P(D)} analysis for the faint end of the count, constraining an amplitude and a slope. A similar approach with the VLA data described here was carried out in \citetalias{Condon12}, where a simple power law was fit to the count below $10 \, \mu$Jy. In this paper we follow the more computationally intensive approach of \citet{Patanchon09} to apply a histogram-fitting procedure for the full range of image source brightnesses. This approach does not require that the source count model be a power law, allowing for more flexibility in accurately modelling the true source count. For completeness we give here a brief summary of the statistics of \textit{P(D)}, providing some specific details on how we applied this to the $3\,$GHz VLA data. For more detailed derivations see \citet{Condon74}, \citet{Takeuchi01}, and \citet{Patanchon09}.  

\subsection{Probability of deflection}
\label{sec:pofd}
The deflection, $D$, at any point (pixel) is an image intensity (in units such as Jy per beam solid angle) at that point. \textit{P(D)} is then the probability distribution of those deflections in some finite region of the image. The differential number count ${dN(S)/dS}$ is the number of sources per steradian with flux densities between $S$ and $S+dS$ per unit flux-density interval. The relative point spread function (PSF) $B(\theta,\phi)$ is the relative gain of the peak-normalised CLEAN beam at the offset of a pixel from the source\footnote{An assumption with the \textit{P(D)} method is that the PSF is constant across the image. With single dish observations or those done at other wavelengths, such as sub-mm or infrared, this may not always be the case. However, with our interferometric image the synthesized beam is set before transformation from the Fourier plane to the image plane. Thus, with our VLA data the PSF is a constant size and shape across the entire image}. The image response to a point source of flux density $S$ at a point in the PSF where the relative gain is $B$ is $x=SB(\theta,\phi)$. The mean number of source responses (e.g. pixel values) per steradian with observed intensities between $x$ and $x+dx$ is $R(x)dx$ \citep[see][for example]{Condon74}, with 
\begin{equation}
\label{eq:rx}
 R\left( x \right) \, dx = 
 \int_{\Omega}\, \frac{dN}{dS} \! \left( \frac{x}{B(\theta,\phi)} \right) 
       \, B(\theta,\phi)^{-1} \, d\Omega\, dx \, .
\end{equation}

The PDF, or probability distribution function, for the observed flux density in each sky area unit (in this case an image pixel) is the convolution of the PDFs for each flux density interval over all flux densities -- this is \textit{P(D)}. The convolution in the image plane is just multiplication in the Fourier plane of the individual characteristic functions. In this case $D$ is the total flux density from all sources with the observed flux density $x$. Thus,  $P(w)$ is
\begin{equation}
\label{eq:pofw}
 p(\omega) = \exp \left[ \int_0^{\infty} R\left(x\right) 
    \exp \left(i \omega x\right) dx - \int_0^{\infty} R\left( x \right) dx 
     \right] \, ,
\end{equation}
and \textit{P(D)} is the inverse Fourier transform of this,
\begin{equation}
\label{eq:pofd1}
 P(D) = {\cal{F}}^{-1} \left[ p(\omega) \right] .
\end{equation}
 
The \textit{P(D)} distribution in a noisy image is the convolution of the noiseless \textit{P(D)} distribution with the noise intensity distribution. Convolution is equivalent to multiplication in the Fourier transform plane, and the Fourier transform of a Gaussian is a Gaussian, so for Gaussian noise with rms $\sigma_{\rm n}$,\footnote{In the case of single dish observations, or steep-slope counts ($\gamma>2$), the mean deflection above absolute zero $\mu$ should also be subtracted off, such that $D$ would then represent the deflection about $\mu$ rather than zero. The mean deflection can be found from $ \mu = \int x Rdx$. The zero point of the \textit{P(D)} distribution is lost in an interferometer image, which has no ``DC" response to isotropic emission, so the zero point must be a free parameter when fitting our VLA data to model \textit{P(D)} distributions.}
\begin{equation}
\label{eq:pofd2}
P(D) = {\cal{F}}^{-1}\left[p(\omega)\exp{\left({-\sigma_{\rm n}^2\omega^2\over 2}\right)}\right].
\end{equation}

The task then boils down to using the measured \textit{P(D)} to constrain a model for $dN/dS$, via $R(x)$, for a given noise and beam.

\subsection{Implementation}
\label{sec:imp}

When calculating the \textit{P(D)} distribution we use very fine binning in flux density: $2^{18}$ bins with bin size $= 0.04 \, \mu$Jy beam$^{-1}$. The output PDF is then interpolated onto the bins used for the image histogram to perform the fit. We calculate and fit \textit{P(D)} over the entire range of pixel values in the given image. For the image histogram we use a bin size of $0.3\, \mu$Jy beam$^{-1}$ below $D = 10 \, \mu$Jy beam$^{-1}$. However, for pixels with flux densities above $10 \, \mu$Jy beam$^{-1}$ there would be very few pixels per bin, because of the small bin size as well as the lack of bright sources in the image; thus a majority would have value 0 or 1. To ensure a large enough number of pixels per bin (to use a Gaussian approximation for fitting) we used expanded bin sizes in the tails. The bin size above $10 \, \mu$Jy increases to ensure a minimum of 10 pixels in all bins. A total of 65 bins were used for the central $5 \,$arcmin region, spanning the range $-7\, \mu$Jy beam$^{-1}$ to $6900\, \mu$Jy beam$^{-1}$. 

\subsubsection{Image noise}
\label{sec:noise}

It is important to have an accurate measure of the instrumental noise in the image for analysis. As mentioned in Section~\ref{sec:obs} we created 16 images from the different S-band frequency sub-bands. The images created from the UV data should have constant instrumental noise across the images, before any primary beam corrections and neglecting any deconvolution artifacts or contamination from dirty-beam sidelobes (which in our case were small with the largest dirty-beam sidelobe contamination being $\simeq0.1 \, \mu$Jy beam$^{1}$). We used the \textsc{AIPS} task IMEAN to calculate the rms noise values of the CLEANed sub-band images in four large areas well outside the primary beam of each. This ensures that the contribution from source signals in these regions is negligible. The 16 images were combined with weights inversely proportional to the sub-band noise to create the $3\,$GHz centre image. The noise was then measured again in several large areas outside the $3\,$GHz primary beam area of the centre image. From these measurements we obtained our noise estimate of $\sigma_{\rm n}=1.012 \pm 0.007 \, \mu$Jy beam$^{-1}$, constant across the image, before the primary beam correction. For more details on the imaging process and noise measurements see section 2.4 of \citetalias{Condon12}.

\begin{figure}
\includegraphics[scale=0.375,natwidth=9in,natheight=9in]{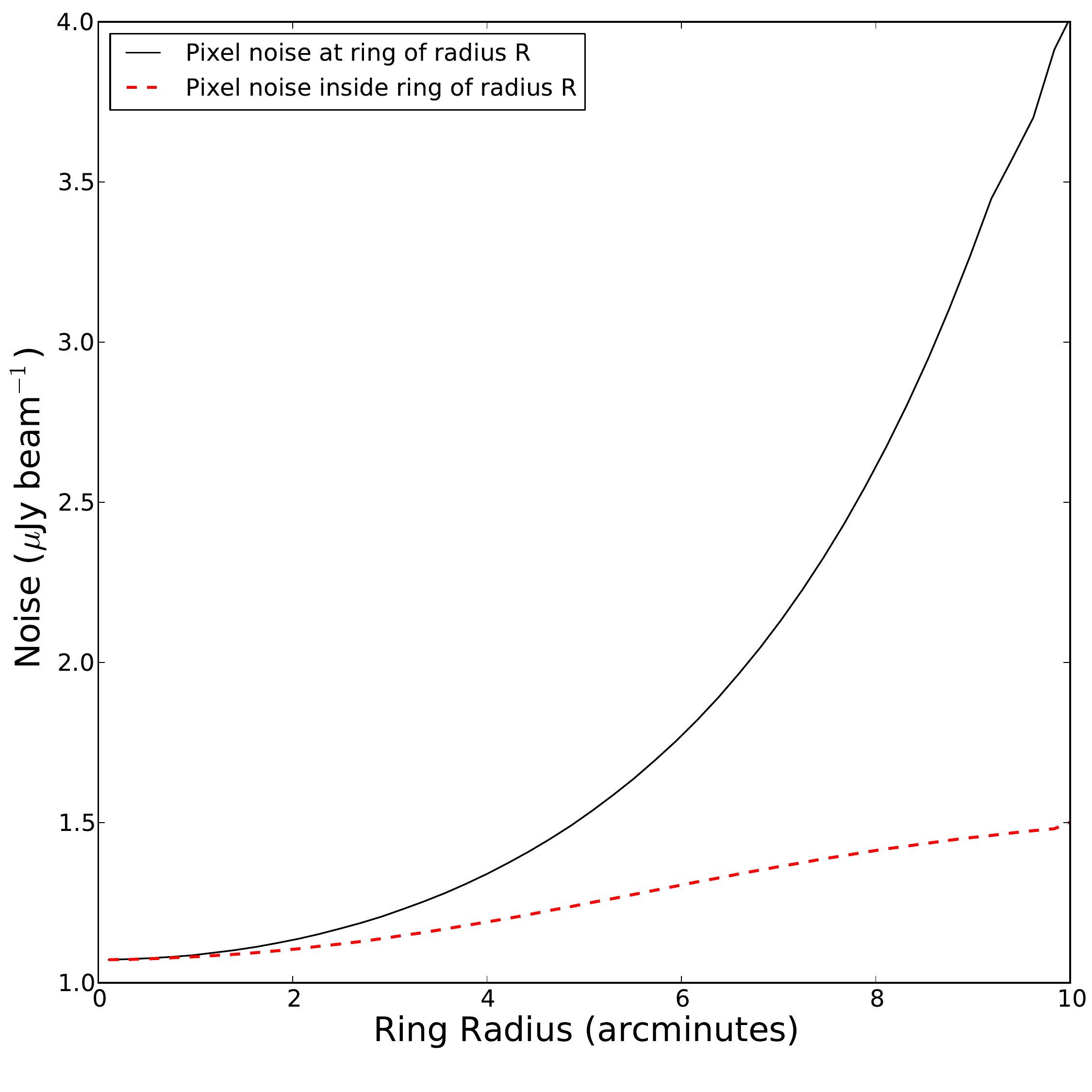}
\caption{ Change in image noise as a function of ring radius. The lines show how the noise at each ring changes with distance from the centre (black solid line) and the weighted noise within a ring of that radius (red dashed line).}
\label{fig:noiseplot}
\end{figure}

For the basic \textit{P(D)} calculation using eq.~(\ref{eq:pofd2}), it is assumed that the noise, $\sigma_{\rm{n}}$, is constant across the image. However, for our VLA data this is not the case. While the true instrumental noise does not change, because of the primary beam correction and frequency weighting effects, the noise measured in $\mu$Jy beam$^{-1}$ increases radially with distance, $\rho$, from the pointing centre. The noise for the ring of pixels at a radius of $5\,$arcmin has already increased from $1.08\, \mu$Jy beam$^{-1}$ to $1.447 \, \mu$Jy beam$^{-1}$. However, the actual noise contributing to the \textit{P(D)} is a weighted combination of the variance of the rings inside some set radius.
Thus, for a circle of radius $5\,$arcmin the weighted effective noise from all the rings inside is $1.255 \, \mu$Jy beam$^{-1}$, as seen as the red dashed line in Fig.~\ref{fig:noiseplot}. We have to choose an area where the variation in the noise is not too large, since for a \textit{P(D)} analysis we want $\sigma_{\rm n}$ to be roughly constant. For highest accuracy we would also like $ \sigma_{\rm n} \leq \sigma_{\rm c}$, where $\sigma_{\rm c}$ is the confusion noise; and yet we want the area to be as large as possible to provide the most samples. We chose to carry out the main \textit{P(D)} calculation within the central 5 arcmin, where the fractional change in the noise has a broad minimum and the effective noise is $\leq \sigma_{\rm c}$.

When binning the pixels for the histogram, weighting must be applied for the histogram to reflect the effective width of $\sigma_{\rm n}^*=1.255 \, \mu$Jy beam$^{-1}$. To accomplish this the area is split into sub-rings with radii (as measured from the mid-point radius of the ring) increasing by $0.11\,$arcmin. A histogram is made for each ring and a value equal to
\begin{equation}
 \label{eq:weight2}
 w_k=\frac{1}{\sigma_{{\rm n}_k}^{4}},
 \end{equation}
gives the pixel weight in the $k$th ring. The $\sigma_{{\rm n}_k}$ is the value of the noise, after the primary beam correction, in the $k$th ring (the black line of Fig.~\ref{fig:noiseplot}).  The weights, $w_k$, go as $\sigma_{{\rm n}_k}^{-4}$ because in this case the estimator is a variance, and thus the weights go as the square of the variance, or the variance of the variance \citepalias[see section 3 of ][for a more detailed discussion of the noise, weighting, and choice of area]{Condon12}. These weights are applied to each ring histogram and the histograms are combined. The rings used for the central $5\,$arcmin can be seen in Fig.~\ref{fig:gridz}. This weighting scheme takes into account the areas of the rings but also favours the more sensitive (lower noise) rings. 

The weighting also affects the uncertainties on the bins for the combined histogram. There are 23 rings in the central area and thus 23 histograms; each of those histogram's bins has Poisson uncertainties of $\varsigma_{i,k}=\sqrt{n_{i,k}}$, for the $i$th bin of the $k$th histogram (or $k$th ring). The uncertainties of the combined histogram are then a weighted combination of these such that,
\begin{equation}
{\varsigma_i}^2=\sum_k n_{i,k}w_k^2,
\label{eq:weight3}
\end{equation}
which can be seen compared with the standard $\sqrt{n_i}$ Poisson value in the bottom panel of Fig.~\ref{fig:corrm}. It is these bin uncertainties that are used when model fitting.

Additionally, to increase the amount of data used to constrain the count we ran the fitting in two other zones. The first extends from $5$ to $7.5\,$arcmin, and the second covers from $7.5$ to $10\,$arcmin. The effective noise inside this second zone is $2.005 \, \mu$Jy beam$^{-1}$ and the effective noise inside the third zone is $3.550 \, \mu$Jy beam$^{-1}$. With just the $0$ to $5\,$arcmin zone we are sampling about $8\,$per cent of the available pixels. The use of all three zones brings that up to around $32\,$per cent of the image pixels. While this still leaves a large fraction of the total image unused for constraining the count, outside a $10\,$arcmin radius the instrumental noise overwhelms the confusion noise. 

\begin{figure}
\includegraphics[scale=0.385,natwidth=9in,natheight=9in]{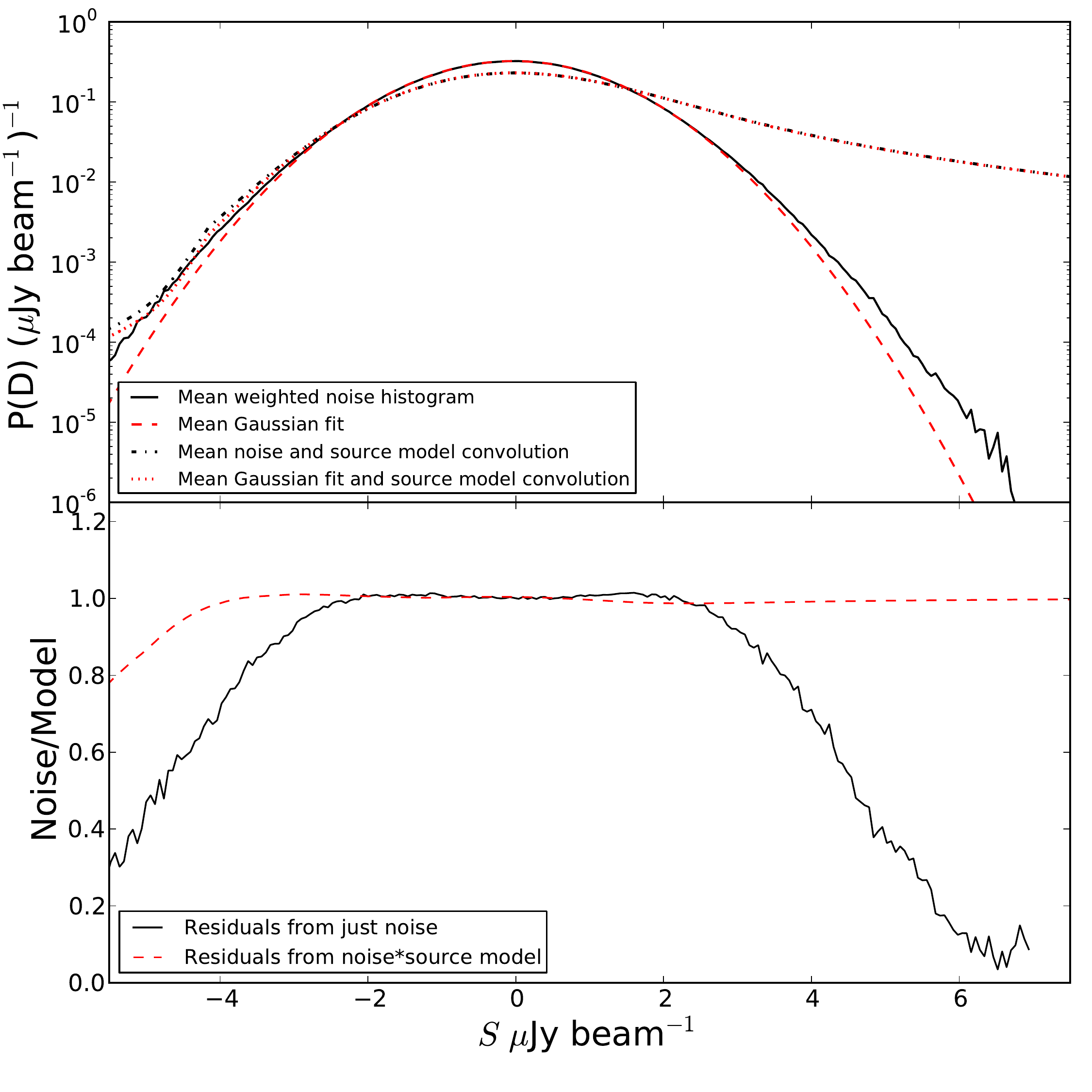}
\caption{Differences in PDFs of purely Gaussian noise and weighted varying noise. The top panel shows the mean results from 100 simulated noise generations. The solid black line is the mean from the weighted noise image histograms, and the red dashed line is the mean from Gaussians fitted to those histograms. The dot-dashed black line line is the mean from convolving the noise image histograms with a noiseless source count \textit{P(D)} and the red dotted line is the mean from convolving the same source count \textit{P(D)} with the fitted Gaussians. The bottom panel shows the ratios of the means. The black line is the mean from the fitted Gaussians divided by noise image histograms and the red dashed line is the mean from the fitted Gaussian convolution divided by the mean from image convolution. This shows how the noise weighting we apply to our data causes it to deviate from purely Gaussian noise.}
\label{fig:gausnoise}
\end{figure}

\begin{figure}
\includegraphics[scale=0.375,natwidth=9in,natheight=17in]{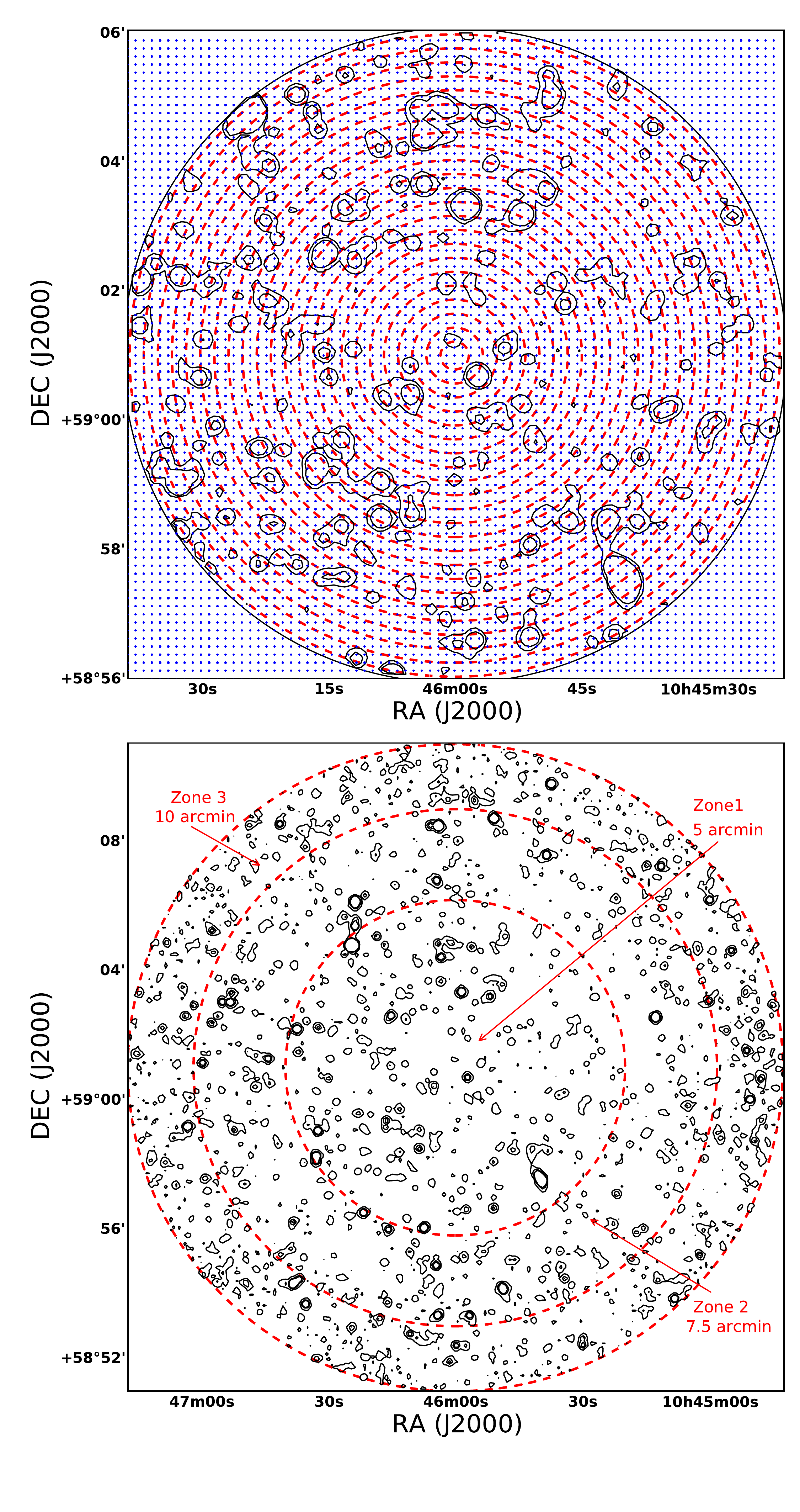}
\caption{VLA $3$-GHz contour images of the Lockman hole. The upper panel shows the central $5\,$arcmin, where the red dashed lines are rings used for weighting the histogram for the primary beam and the blue crosses are the pixel locations from one of the grids, with spacing between the points equal to the beam FWHM. The lower panel is the same image out to $10 \,$arcmin, with the red dashed lines now showing the separation of the three noise zones discussed in Section~\ref{sec:noise}. }
\label{fig:gridz}
\end{figure}

\begin{figure}
\includegraphics[scale=0.375,natwidth=9in,natheight=9in]{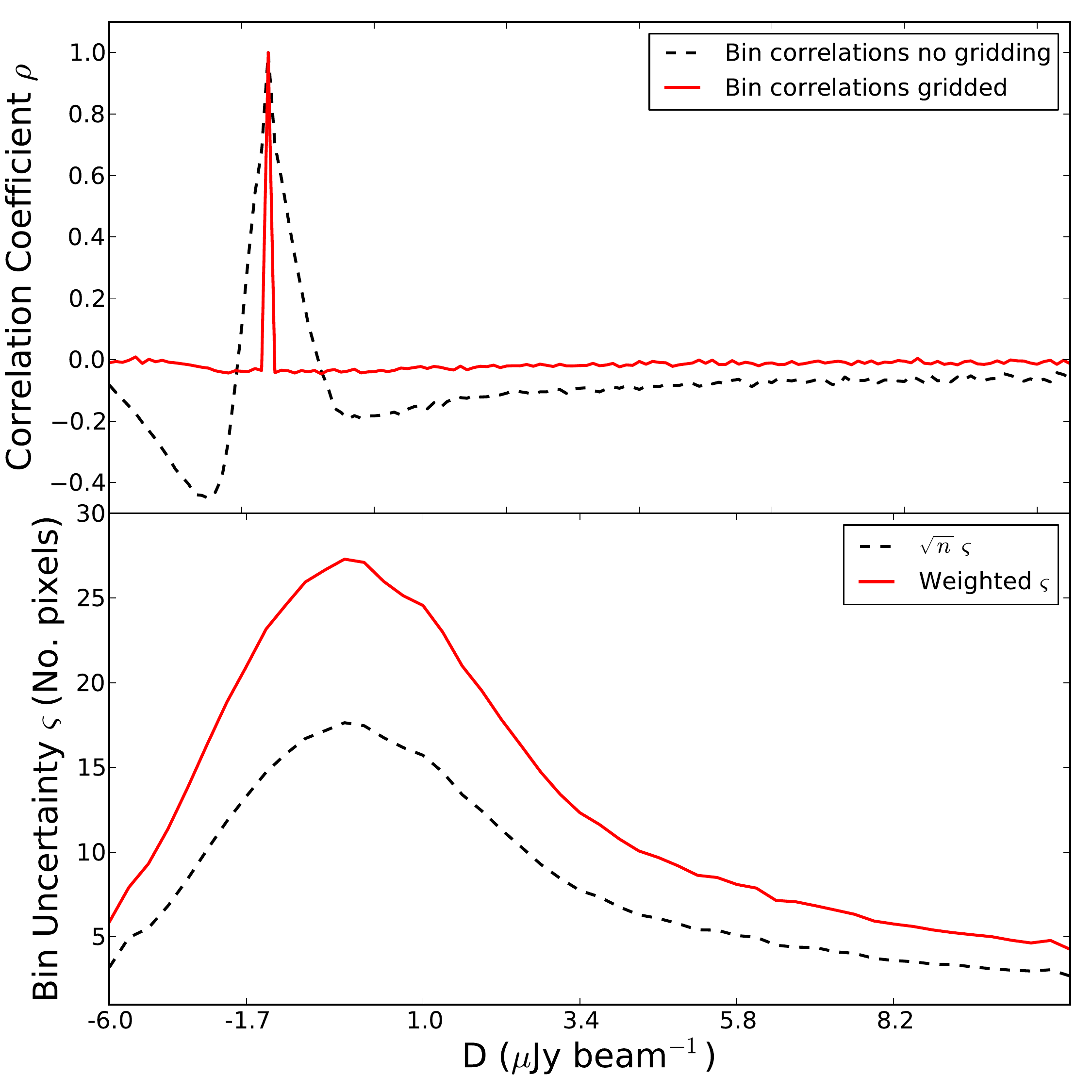}
\caption{Bin correlations and uncertainties. The top panel shows the bin-to-bin correlation coefficients for one row of the correlation matrix at the peak, computed using eq.~(\ref{eq:correlm}). The black dashed line shows the correlation values computed from full pixel histograms of $20{,}000$ simulated images. The red solid line indicates the correlation values for the same bin computed from 20,000 simulated images but with the histograms made from pixels separated by one beam FWHM. The bottom panel shows the uncertainties for each bin. The black dashed line is $\varsigma=\sqrt{n_i}$ and the red solid line is the uncertainty $\varsigma$ due to weighting calculated from eq.~(\ref{eq:weight3}).  }
\label{fig:corrm}
\end{figure}

The frequency-dependent primary beam correction, and our weighting scheme, does mean that our image noise is not purely Gaussian, as is assumed in the \textit{P(D)} calculation. We ran a simulation to see by how much our noise might be deviating from Gaussian and whether this could impact our fitting. We created images of random Gaussian noise, of the same size as our central $5\,$arcmin and convolved them with a Gaussian the same shape and size as our beam. We then applied the same corrections as to our actual data and created a weighted histogram of each using the process described above. A Gaussian was fit and calculated for each sample noise image, and then the noise image \textit{P(D)} and that of the fitted Gaussian were both convolved with a noiseless \textit{P(D)} from a source count model (the specific source PDF used can be seen in Fig.~\ref{fig:pdflog}). After $100$ trials we calculated the mean \textit{P(D)} from the noise histograms, fitted Gaussians, noise histograms convolved with the source model PDF, and fitted Gaussians convolved with the source model PDF. These four means can be seen in the top panel of Fig.~\ref{fig:gausnoise}, with the ratio of the fitted Gaussians to the noise images shown in the bottom panel. We can see that for the noise alone the true weighted histograms do deviate from Gaussians starting at around $3\sigma$, with the largest deviations being about a factor of $2.5$ in the $5\sigma$ region. However, once convolved with the source count \textit{P(D)} the deviation is much smaller. There is then no discernible difference in the two distributions on the positive side. On the negative side the maximum deviation from the Gaussian model is only about a factor of 1.25, and this is only in the $4{-}5\sigma$ range, where in the images there are likely only $0{-}3$ pixels/bin. Thus, with our current data this should not present any bias in the fitting.

\subsubsection{Model fitting}
\label{sec:mcmcfit}
We have developed a code to fit \textit{P(D)} based on a set of input model parameters. The input model need not be a simple power law, and may take on various forms, as long as it is continuous over the chosen flux density range. To fit the parameters we forward-model and use Markov chain Monte Carlo (MCMC) sampling methods. We make use of the publicly available MCMC package \textsc{CosmoMC} \citep{Lewis02}\footnote{\url{http://cosmologist.info/cosmomc/}}, which, while developed for use in cosmological modelling, may be used as a generic sampler, if one provides data, model, and likelihood function. The MCMC code varies the input parameters in order to minimize the chosen fit statistic. Once the chain has past the ``burn-in" phase it converges near the minimum and will then sample the parameter space, drawing from the parameter's proposal density to decide on the next step in the chain. A well chosen proposal density can improve the efficiency of the fitting procedure. For all of our chains we first ran sample chains, with about an order of magnitude fewer steps than the final chains, and used these to compute the covariance matrix of the parameters, which we then supplied to the MCMC code to use for the proposal density.

There has been discussion about the optimal choice of statistic to use for \textit{P(D)} fitting. One possibility is to use the classical $\chi^2$, as done by \citet{Friedmann04} and \citet{Maloney05}. However, the weighting of $1/n_i$, with $n_i$ being the number of pixels in the $i$th bin, will tend to over-weight the bins when $n_i$ is small, giving more weight to the tails of the distribution Since for small numbers the uncertainty is not well modelled by $\sqrt{n_i}$, this option is not ideal. Another choice is to minimize the more correctly calculated negative log likelihood, as done by \citet{Patanchon09} and \citet{Glenn10}. While this method gives proper weighting, the problems come when trying to interpret the goodness of the fit. For the \textit{P(D)} model with Poisson statistics the log likelihood is defined as 
\begin{equation}
 \log \mathcal{L} = - \sum_i n_i \log(p_i) - \log(N!)+\sum_i\log(n_i!).
\label{eq:logl}
\end{equation}
Here $N$ is the total number of pixels in the image, $p_i$ is the probability in the $i$th bin when the PDF is normalised to sum to one, and $n_i$ is the the number of image pixels in the $i$th bin.  In the limit that $n_i\gg1$ this approximates a $\chi^2$ distribution:
\begin{equation}
\frac{\chi^2}{2}\simeq \frac{1}{2}  \sum_i \frac{\left(n_i -Np_i\right)^2}{Np_i}+K,
\label{eq:chi}
\end{equation}
where $K$ is a normalisation factor usually taken to be $K=(1/2)\sum_i(Np_i)$. However, when the log  likelihood of eq.~(\ref{eq:logl}) does not equal the left hand side of eq.~(\ref{eq:chi}) it can be difficult to determine $K$ and therefore difficult to interpret the log likelihood. 

Neither of these two methods takes into account the fact that the pixels in the map (and hence bins in the histogram) are correlated. Due to the sources and noise being convolved with the beam, values in one location will affect neighbouring pixel values within an area roughly equal to the size of the beam \citepalias[or the size of the beam area divided by 2 in the case of the noise; for further explanation see][]{Condon12}. Furthermore, one source, when convolved with the beam, will contribute pixels to multiple bins. Ignoring these issues will underestimate the uncertainties of the bins and correspondingly the uncertainties of the fit parameters. \footnote{Both \citet{Patanchon09} and \citet{Glenn10} discuss the related issue of the optimal smoothing kernel for obtaining maximum signal-to-noise ratio for using \textit{P(D)} to constrain counts. However, it is important to realise that the situation in interferometry is fundamentally different than for single dish data. In direct imaging observations the instrumental noise is (ideally) independent at the map level, and hence it makes sense to further smooth this by a kernel of approximately the beam size \citep[as shown in figure 3 of][]{Patanchon09}. However, for interferometric imaging, the noise is independent in the Fourier plane, and when going to the image plane has \textit{already} been convolved by the synthesized (dirty) beam. \citet{Chapin11} found from simulations of sub-mm data that in the very confused regime the optimal filter is the inverse of the PSF in Fourier space, i.e., the map is de-convolved by the beam; and in the regime dominated by instrument noise the optimal filter is the PSF. In our case, with $\sigma_{\rm n} \simeq \sigma_{\rm c}$ our current weighting scheme may not be optimal for \textit{P(D)}, but is likely close. To determine the ideal weighting and filtering scheme for our type of data would require a more thorough analysis, starting in the Fourier plane and looking at ways to optimize before transformation to the image plane, rather than applying filters post transformation. This is beyond the scope of this paper.}

When dealing with correlated variables the ideal solution is to use the generalised form of $\chi^2$, which includes the covariance matrix of the data. However, tests run using simulated images show this correlation matrix to be highly dependent on the underlying source count model. As we do not know in advance the true source count for our data, we want to avoid biasing the results by using a covariance matrix calculated from simulations performed with only an approximated source count.

In order to remove the bin-to-bin correlations we instead sampled the image using a grid of positions with spacings of one beam FWHM. This should ensure that the pixels are approximately independent and correspondingly that the histogram bins are also independent. Of course the optimal sampling will be a compromise between reducing the correlations, and not losing too much fine-scale information, so it is certainly necessary to test that sampling with $1\times$FWHM spacing is close to the best choice. 

We tested the effectiveness of this method using simulations. We used a simple broken power-law source count of slope $-1.7$ for flux densities less than $10 \,\mu$Jy and $-2.3$ for sources brighter than $10 \, \mu$Jy,  to generate sources that were randomly placed in an image with the same number of pixels as our data image. We convolved these sources with a beam of the same size as ours, and added them to beam-convolved Gaussian noise with $\sigma =1.255 \times 10^{-6}$. We simulated $20{,}000$ realisations in this way, made full histograms of each and also created histograms using pixels sampled from a grid with spacings of FWHM/$\sqrt{2}$, $1\times$FWHM, and $\sqrt{2}\times$FWHM. We computed the mean number of pixels per bin from these and then computed the corresponding correlation matrix. Each entry in the correlation matrix was computed such that,
\begin{equation}
\rho_{i,j}={1\over 20000}\sum_{k}\frac{(n_{i,k}-\mu_i)(n_{j,k}-\mu_j)}{\varsigma_i\varsigma_j},
\label{eq:correlm}
\end{equation}
the diagonals of which are equal to 1. The correlation coefficient, $\rho_{i,j}$, is equal to $C_{i,j}/\varsigma_i\varsigma_j$, where $C_{i,j}$ is the covariance of the $i$th and $j$th bin. One row from this matrix, near the peak of the histogram, is plotted in the top panel of Fig.~\ref{fig:corrm}. This shows that by taking FWHM-separated samples from the grid we remove nearly all of the correlation between the bins; the off-diagonals of the gridded simulation are all zero within statistical error. The samples with grid spacings of FWHM/$\sqrt{2}$ showed higher off-diagonal correlations. The $\sqrt{2}\times$FWHM samples have roughly the same correlations as using $1\times$FWHM, but with lower resolution. Thus we chose to use grids with the FWHM spacing. 

The images we use have a beam width of approximately six pixels. Hence the FWHM grid which samples the image could be shifted in RA and Dec, with 36 different choices possible without repeating any pixels. An example of the grids can be seen in the top panel of Fig.~\ref{fig:gridz}, where the blue mini-crosses represent the positions of the image pixels selected for binning that grid. From the 36 histograms we are able to compute the scatter for each bin, which can be used as a check on the calculated bin uncertainties described in eq.~(\ref{eq:weight3}).

We chose to carry out the MCMC fitting by minimising 
\begin{equation}
\chi^2= \frac{1}{2}  \sum_i \frac{\left(n_i -Np_i\right)^2}{\varsigma_i^2},
\label{eq:leasts}
\end{equation}
where the uncertainties used were not the usual Poisson $\sqrt{n_i}$ error bars, but rather (due to weighting effects from the primary beam) those from eq.~(\ref{eq:weight3}). We performed MCMC trials on our VLA data using both eq.~(\ref{eq:leasts}) and eq.~(\ref{eq:logl}) (both using the gridded image histograms). Comparisons of the output fit parameters for the different methods can be seen in Fig.~\ref{fig:compare}. Although the outputs from the two methods are consistent, because the value of the log likelihood does not equal the $\chi^2/2$ it is difficult to interpret the goodness of the fit. 

Also in  Fig.~\ref{fig:compare} we show the gridded method against the results of a trial using all of the image pixels. The output is not significantly different for the parameters; however, as mentioned, the full resolution method underestimates the limits, the $68\,$per cent error region being roughly a factor of 1.5 to 3 times smaller in $\log_{10}dN/dS$. We know that the fits performed with the gridded data use approximately independent samples. Even though we do lose some resolution we believe this method to be more statistically robust and to model more accurately the variance and correlations. 

\section{Choice of Model}
\label{sec:model}
In \citetalias{Condon12} a single power-law model was fit to the data in this field. The best fitting single power law in the range $ 1 < S <10 \, \mu$Jy was ${dN/dS} = 9000S^{-1.7}\,$Jy$^{-1}$ sr$^{-1}$. It was noted that power law models from \citet{Condon84b},  ${dN/dS} = 9.17 \times 10^4S^{-1.5}\,$Jy$^{-1}$ sr$^{-1}$, and \citet{Wilman08} simulations, ${dN/dS} = 2.5 \times 10^4S^{-1.6}\,$Jy$^{-1}$ sr$^{-1}$, were both reasonably good approximations to the data in this range (assuming $\langle\alpha\rangle=-0.7$ to convert from $1.4 \,$GHz to $3 \,$GHz).  However, it is the case that no single power law fits well across the whole $\mu$Jy region. 

\subsection{Modified power law}
\label{sec:mplmodell}
Since the single power-law model had already been explored, we first decided to try fitting a modified power law of the form
\begin{equation}
\frac{dN}{dS} = \kappa S^{\alpha+\beta\log_{10}S+\gamma\left(\log_{10}S\right)^2} ,
\label{eq:powerl}
\end{equation}
in the range $ 0.01 < S <60 \, \mu$Jy. For $S > 60 \, \mu$Jy we connected the modified power law to the model from \citet{Condon84b} (scaled to $3 \,$GHz using $\langle\alpha\rangle=-0.7$), where this model is in good agreement with known counts. We chose the cut-off at $60\, \mu$Jy so that we would fit the data not just in the $\mu$Jy region but also in the slightly brighter area where the count from \citet{Owen08} was found to be higher than expected at $1.4 \,$GHz. We fit for $\alpha, \beta,$ and $\gamma$, while $\kappa$ was calculated as a normalisation constant to ensure continuity at $S=60\, \mu$Jy. The results are presented in Section~\ref{sec:results}.

While the modified power law is a better fit than a simple power law, one would still like to be able to constrain the shape of the count in more detail over different intervals of flux density. With the modified power law, the fit parameters are not very sensitive to the region $S \leq \sigma_{\rm n}$, even though there is still information in the image at these faint flux densities. This model also does not allow us to investigate the faintest limits for which constraints are still possible. Therefore, we have followed the approach of \citet{Patanchon09} and \citet{Glenn10} and fit a phenomenological parametric model of multiple joined power laws, allowing for more variation in the shape of the count. In this approach we fix the position in $\log_{10}(S)$ of a fixed number of nodes, and fit for the node amplitude of log$_{10}$ ${dN/dS}$. Between the nodes the count is interpolated in $\log$ space to ensure a continuous function, with the count outside the highest and lowest nodes set to zero. The node amplitudes do not actually represent the value of ${dN/dS}$ at the positions of the nodes, but rather represent an integral constraint on some region surrounding the node. Therefore, the best-fit position of any given node depends not only on the underlying source count but also on the number, or spacing, of the nodes, and also the type of interpolation used between the nodes.

\subsection{Node-based model}
\label{sec:nodem}
\begin{figure*}
\includegraphics[scale=.6,natwidth=12in,natheight=12in]{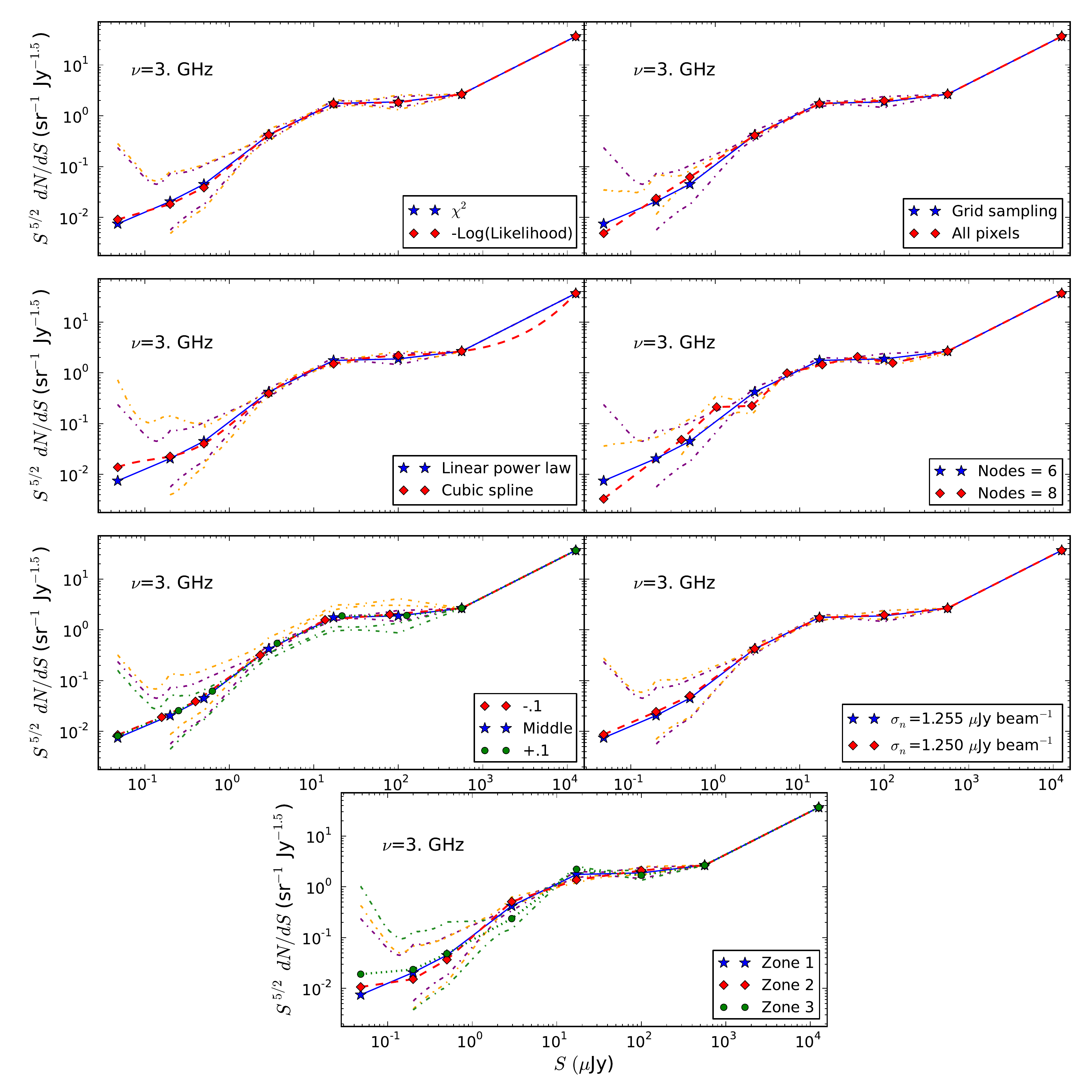}
\caption{ Comparison of MCMC output for $3\,$GHz source count with different model settings. Points and corresponding lines are the means of each parameter's marginalised probability distribution and dot-dashed lines represent the $68\,$per cent confidence regions. The blue solid line is the same in all six plots and represents the marginalised means for Zone 1 reported in Section~\ref{sec:results}. Top left panel: blue solid line is the MCMC run with minimising $\chi^2/2$, the red dashed line is for fits minimising the log likelihood. Top right panel: blue solid line is run using the gridded pixel histograms, and the red dashed line is with a histogram using all the image pixels. 2nd row left panel: blue solid line run using linear interpolation in $\log_{10}(S)$ between the nodes, while the red dashed line uses a cubic spline model. 2nd row right panel: blue solid line is for a run fitting six nodes and red dashed line is with eight. 3rd row left panel: blue solid line is the run the with initial positions in flux density, while the red dashed line is the run with node positions set as $\log_{10}(S_{\rm{centre}}) -0.1$ and the green dotted line is the run with $\log_{10}(S_{\rm{centre}})+0.1$.  3rd row right panel: blue solid line is the run with a fixed noise of 1.255 $\mu$Jy beam$^{-1}$ and the red dashed line allows the noise to float, with marginalised mean $\sigma_{\rm n}^*=1.250$. Bottom panel: blue solid line is the run using the noise zone 1, while the red dashed line is for noise zone 2, and the green dotted line is for noise zone 3.}
\label{fig:compare}
\end{figure*}

The choice of the number and position of the nodes is somewhat subjective. There need to be enough nodes across the flux-density range to be able to account for changes in the underlying count, and the choice is also influenced by the resulting uncertainties on the parameters. The fits of the node positions are degenerate; neighbouring nodes will be most strongly correlated, and so, adding too many nodes will increase the correlations and parameter degeneracies. We examined trials using five, six, and eight nodes. We found that there was no significant change in the $\Delta \chi^2$ with the total of eight nodes, and with six the results were most consistent over repeated trials. Comparison of the results with different number of nodes can be seen in Fig.~\ref{fig:compare}; based on this, we decided to fit six nodes, spaced roughly evenly in $\log_{10}S$. The value of the faintest node is to be considered only as an upper limit, since the code cannot distinguish between low amplitude values and zero. Therefore, the situation is effectively that we fit five well constrained nodes and one upper limit. In the \textit{P(D)} calculation we also considered two additional brighter nodes at fixed ${dN/dS}$ values. The highest node is far above any source in our field, and it was found that changing its value during \textit{P(D)} calculation had no effect on the output. The second highest is also in a very sparsely populated flux density area for our image (only one source brighter). These two node positions are in a well-constrained  range of the $1.4\,$GHz source count, so rather than fitting for these nodes their values were estimated from existing $1.4\,$GHz source counts, scaled to $3 \,$GHz using $\langle\alpha\rangle =-0.7$.  Adding these extra nodes is essentially the same as adopting a prior on the brightest count region considered.

The positions for the six nodes were chosen through trial and error.  We found that the results were not sensitive to a faintest node below $-7.3$, in $\log_{10}(S)$, and thus this position was chosen for the lowest node. For the second faintest node, we found that any nodes placed in the region between the faintest and $\sim 0.25\sigma_{\rm{n}}$ were difficult to constrain and very degenerate for more than one in that region. We therefore chose to place the second node at about a quarter of the instrumental noise, which produces reasonably robust constraints. As far as the spacing between the second and sixth nodes, the requirements are to have fairly evenly spaced nodes in $\log_{10}(S)$, while still having at least one node in the $\mu$Jy region, one near the Owen $\&$ Morrison (2008) flux density limit, and one between that and the fixed node near our brightest flux density. We ended up with four nodes (three power laws) encompassing the region from $0.2$ to $17.2\, \mu$Jy, fully covering the region fit in \citetalias{Condon12} and the Owen $\&$ Morrison (2008) sources. Although the node placement was fixed, to make sure that the precise positions did not bias the results we also ran chains at $\pm 0.1$ in $\log_{10}S$ of the centre nodes, the results of which can also be seen in Fig.~\ref{fig:compare}. Since no discernible difference was observed when varying the positions, for the rest of the analysis the centre positions were adopted. 

Since the source count comes from a redshift integral over luminosities, the count must be continuous between $S_{\rm{min}}$ and $S_{\rm{max}}$. $S_{\rm{max}}$ is set by the flux density of the brightest node, $0.0126\,$Jy, which, as above, was chosen to be brighter than any source in our image, but not so bright as to greatly increase the range (so that our bin size could be kept as small as possible). $S_{\rm{min}}$ in our case is set by the number of bins, and is thus $0.0126/2^{18}=0.04 \, \mu$Jy. Since we are fitting for nodes at only a few positions, it is necessary to interpolate the count between the nodes. As well as using linear interpolation (multiple power laws), we considered a cubic spline model, with the cubic spline interpolation done in $\log_{10}dN/dS$ and $\log_{10}S$. We ran chains using both models while keeping other variables fixed, and compared the output, which can be seen in Fig.~\ref{fig:compare}. The comparison is not straightforward, since the values at each node do not have exactly the same meaning, being effectively integral constraints over different flux density regions. However, the two methods produce very similar results: the marginalised means are almost exactly the same, but the uncertainties in the fainter regions are larger for the cubic spline model. For simplicity we decided to use the power-law model for the rest of the analysis.

Some additional constraints on the fitting parameters were applied to ensure physically reasonable results. A prior on the background temperature from the integrated count was used. It was set as a cut-off, such that any count model yielding a temperature greater than $95\,$mK at $3 \,$GHz was not considered. This was imposed to allow the count to produce (but not overproduce) the background temperature seen by ARCADE 2 of around $70$ mK. This is a very weak prior, and hence very reasonable to impose, as it not only exceeds the ARCADE 2 value but also greatly exceeds previous source count temperature estimates of $13\,$mK. It is important to set some limit on the amplitude of the faintest nodes, where the data constraints are weakest. For the brighter nodes, a starting estimate of the count was given by approximating known source counts around the node at $1.4\,$GHz scaled to $3\,$GHz. High and low cut-offs were placed on the nodes, limiting the region to be sampled. These were chosen based on the observed high and low count values in the region around the node measured using the compilation of $1.4 \,$GHz source counts from \citet{dezotti09}, scaled to $3\,$GHz. For the nodes fainter than the current cut-off as set by \citet{Owen08}, starting estimates were based on the scaled Condon (1984) model at $1.4\,$GHz. Limits were placed on the sampling space by extrapolating two lines (in log-log space) from the current cut-off, one with a positive slope and one a negative slope. The extreme allowed values for the last node, at $\sim 0.05\, \mu$Jy, were $20$ and $14$ (in $\log_{10}[dN/dS]$). This yielded a wide area to be sampled in a region where no previous information existed. 

It is very important to have an accurate value for the instrumental noise in this calculation, because it convolves the noise-free \textit{P(D)} distribution. Unless  $\sigma_{\rm n} \ll \sigma_{\rm c}$, then small changes in $\sigma_{\rm n}$ can have a significant effect on the output, particularly in the faint flux density regime. Our estimate for the confusion noise is roughly the same as our estimate of the effective instrumental noise inside the $5\,$arcmin ring, $\sigma_{\rm c} =1.2 \simeq \sigma_{\rm n}^{*}=1.255$. Since our noise estimate comes from a weighted average of the instrumental noise of the 16 frequency sub-band images, and then a weighted average of the noise after primary beam correction, any errors in the measurement or calculation of those would affect our calculated noise value. To allow for the possibility of uncertainty in our noise value we performed the MCMC \textit{P(D)} fitting with: (1) the noise fixed at the calculated values for $\sigma_{\rm n}^*$ for each model; and (2) allowing the noise to be a free parameter. In this latter case the calculated noise value was given as a starting estimate for the fitting and we allowed a sampling range of $(1.255\pm 0.05) \, \mu$Jy beam$^{-1}$. 

In the modified power-law case the marginalised mean for the noise is $\sigma_{\rm n}^*=1.268 \pm 0.005$, while the node-based model gives a marginalised value of $\sigma_{\rm n}^*=1.250 \pm 0.006$. These are consistent with the original estimate of $1.255\, \mu$Jy beam$^{-1}$. The results of fitting with the noise being variable versus fixed can be seen in the bottom left panel of Fig.~\ref{fig:compare}. The noise parameter is strongly degenerate with the faintest two node amplitudes. These nodes do not contribute much to the bright tail of the \textit{P(D)}, but mainly affect its width. This explains why, for the variable noise case, the faintest two nodes are slightly higher than in the fixed noise case, since the fitted $\sigma_{\rm n}^*$ is smaller. For both models the fixed and variable noise results are consistent within uncertainties. For the rest of the analysis only the fixed noise results are used.

In terms of the multiple noise zones, the three zones were all fit independently; the results are shown in Fig.~\ref{fig:compare}. We also fit all three zones simultaneously, such that the fit $\chi^2$ was a sum of the individual $\chi^2$s. So in this case we minimized
\begin{equation}
\chi^2_{\rm total}=\sum_i \chi^2_{i},
\end{equation}
where $\chi^2_{i}$ is the $\chi^2$ of eq.~(\ref{eq:leasts}) from each zone for a given set of input model parameters. The results presented in Section~\ref{sec:results} report the fitting of just the first zone (with the lowest noise) and the three zones together ,for both the modified power-law model and the node-based model.

\section{Simulations}
\label{sec:sim}
\subsection{SKADS simulated image}
\label{sec:skads}
To test our model and statistical approach we used data from the SKADS (Square Kilometre Array Design Studies) SKA Simulated Skies (S$^{3}$) simulation \citep{Wilman08}. The S$^3$ simulation is a large-scale semi-empirical model of the extragalactic radio continuum sky at several frequencies. The simulated sources were drawn from calculated luminosity functions, with evolving bias factors $b(z)$ for different galaxy types (radio-loud AGN, radio-quiet AGN, and star-forming galaxies). These were inserted into an evolving dark matter density field. This simulation, therefore, has realistic approximations of the known source counts and contains both small and large-scale clustering. Using these data allowed us to test not only the functionality and accuracy of our code, but also any effects that small-scale (beam-sized) clustering might have on the output, by comparing the fitted model to the known input. 

We used the simulated data at $1.4 \,$GHz, the closest frequency in the simulation to our VLA data. The full size of the simulation is $400\,$deg$^{2}$, from which we extracted the central $1\,$deg$^{2}$. The simulated image was constructed to have the same beam and pixel size as our VLA data. Random (beam-convolved) Gaussian noise was added to the simulated image, with $\sigma_{\rm{n}}=2.14 \, \mu$Jy beam$^{-1}$ rms. This noise value is slightly larger than that of our VLA image central $5\,$arcmin due to the simulation being at $1.4\,$GHz instead of $3\,$GHz. The model count was set up as described in Section~\ref{sec:model}, with six variable nodes and two fixed ones. The faintest node was set at $10\,$nJy, as this was the faintest flux density simulated in the data. The second node was set at $0.1\sigma_{\rm{n}}$. 

\begin{figure}
\includegraphics[scale=0.375,natwidth=9in,natheight=9in]{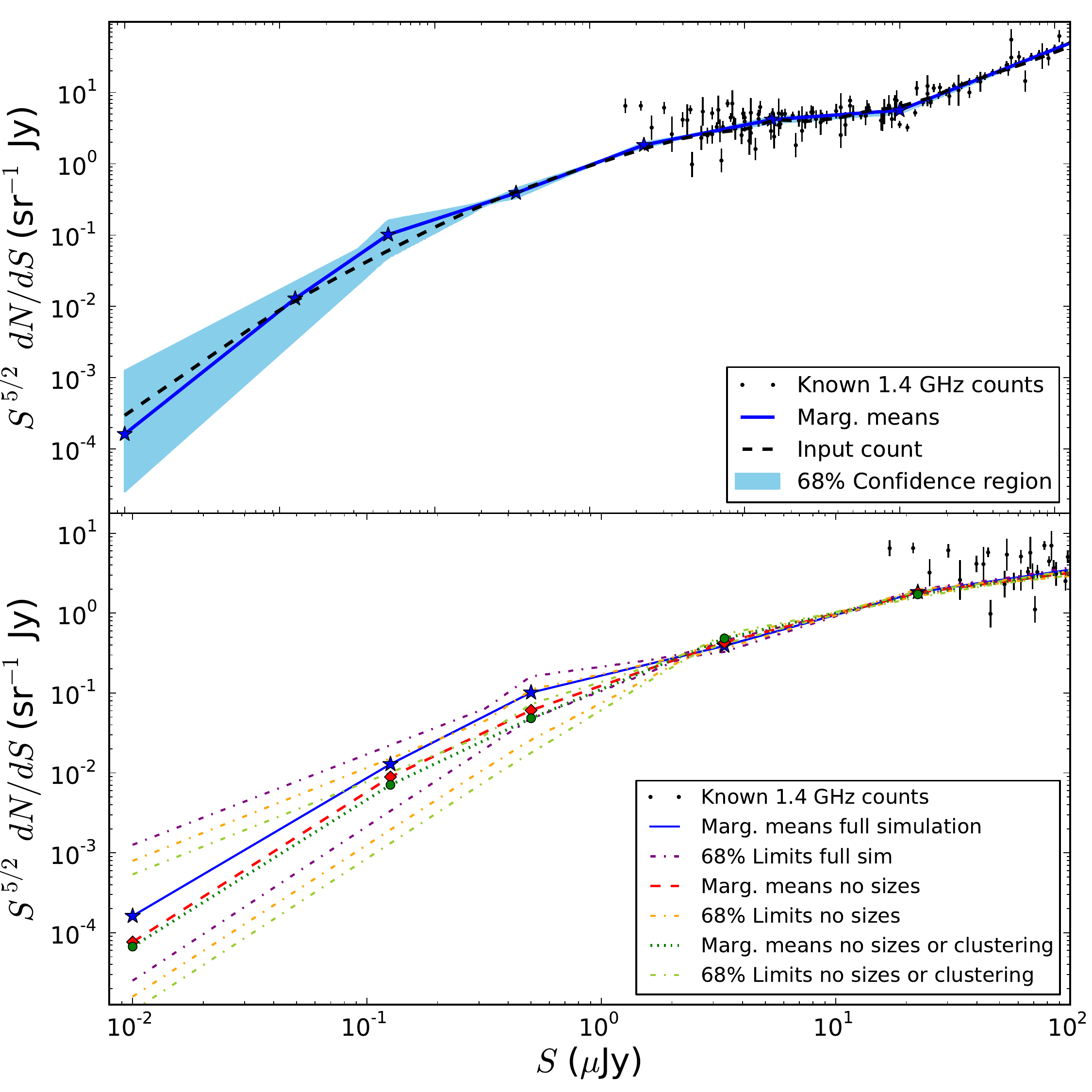}
\caption{Euclidean-normalised source count from SKADS simulation. The black points are real data counts from the \citet{dezotti09} compilation. Top: MCMC marginalised mean (blue line and points) node positions. The black dashed line is the input source count model from \citet{Wilman08}. The shaded area is the $68\,$per cent confidence region. Bottom: The marginalised mean node positions from the simulated image, zoomed in on the region $S\le 2 \, \mu$Jy, taking into account source sizes and clustering (blue solid line), but with sources all unresolved (red dashed line), and for unresolved sizes and random positions (green dotted line), with $68\,$per cent confidence regions as the dot-dashed purple, orange, and green lines.  }
\label{fig:sim_counts}
\end{figure}

\begin{figure}
\includegraphics[scale=0.375,natwidth=9in,natheight=9in]{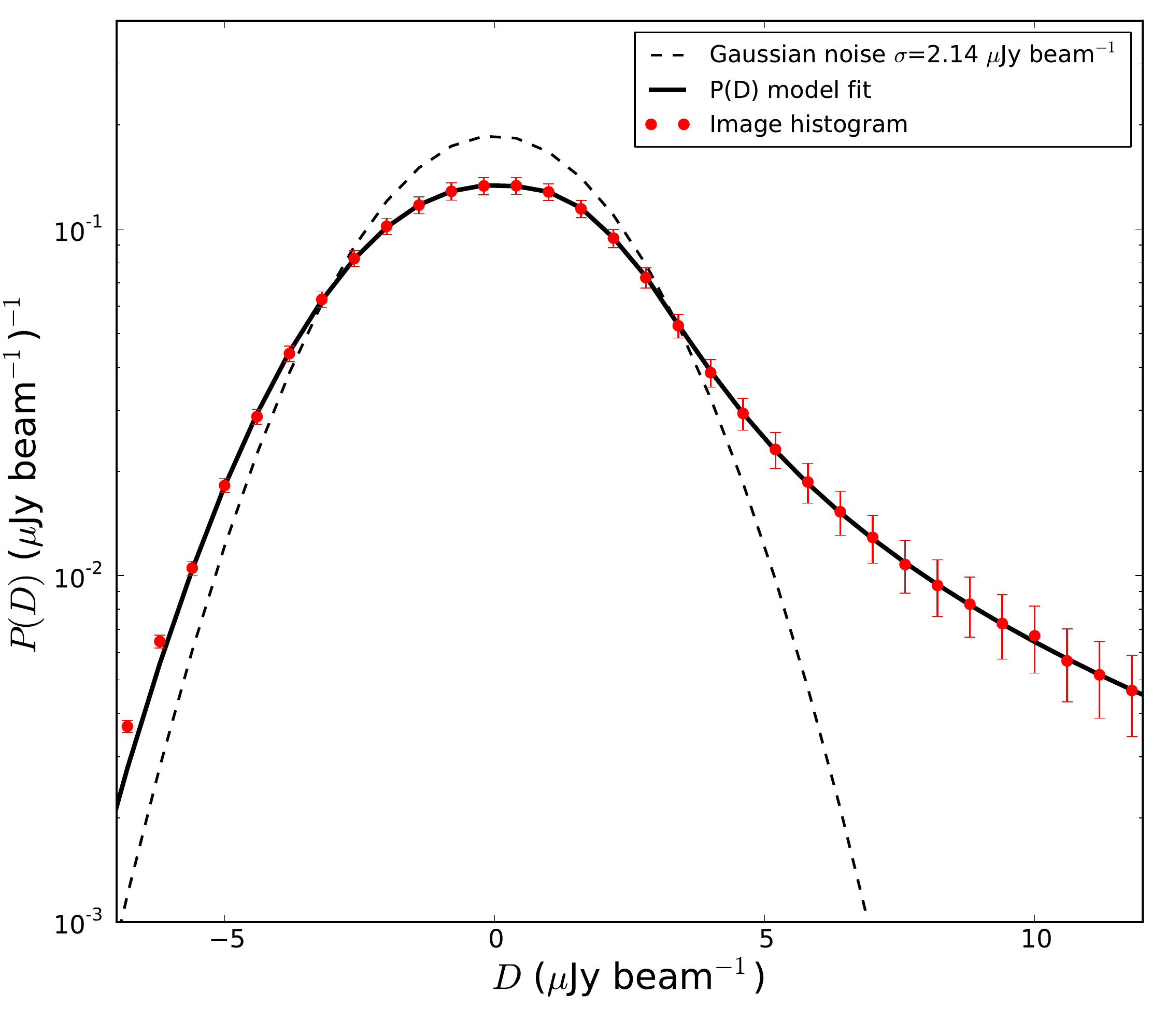}
\caption{ Comparison of pixel histograms from $1.4 \,$GHz simulation (red dots) with the output of the marginalised means model \textit{P(D)} using as input six variable nodes (solid line) and a fixed $\sigma_{\rm{n}} = 2.14 \, \mu$Jy beam$^{-1}$ noise level (dashed line).}
\label{fig:pdflog14s}
\end{figure}

The output from the MCMC fitting to the simulated data can be seen in the top panel of Fig.~\ref{fig:sim_counts} and the \textit{P(D)} distributions are shown in Fig.~\ref{fig:pdflog14s}. The plot shows the marginalised mean amplitudes from each parameter's likelihood distribution for all six nodes. The values for the six nodes and the  $\chi^2$ values at each point in the chain can be used to compute $68\,$per cent confidence intervals (useful for examining the full likelihood surface, since there are shape changes due to the parameter degeneracies, as discussed in Section~\ref{sec:degen}). The results from this simulated image indicate that our fitting procedure is unbiased; the input source count model is always within the relevant confidence regions. There is some slight deviation from the input count for the faintest three nodes. This is perhaps not  unexpected, since this region is well below the instrumental noise. However, while the error bars on the faintest node are large, it is important to note that that the count is still constrained; even the $95\,$per cent limits for this node do not reach the high and low limits given to the MCMC routine. The marginalised mean for the faintest node is within $1\,$per cent of the input value, even though this is two orders of magnitude below the noise limit. This test shows that the method and model are not only capable of fitting the underlying source count of an image, but that there is still information about the count well below the instrumental noise, as long as that noise value is known well.

\subsection{Clustering and source sizes}
\label{sec:cluster}
The simulated image included sources with varying sizes, sources with multiple components, and the underlying clustering information. In \citet{Wilman08} the angular two-point correlation function, $w(\theta)$, is shown for the full simulation; this is a little higher than measurements made by \citet{Blake04} at a somewhat brighter flux density limit. We computed $w(\theta)$ for the specific $1\,$deg$^2$ simulated sample. The function $w(\theta)$ is usually approximated by a power law of the form $w(\theta)=A\theta^{-\alpha}$( or sometimes written as $w(\theta)=(\theta/\theta_\phi)^{-\alpha}$). \citet{Blake02a,Blake02b} found $A =1.0\times10^{-3}$ and $\alpha=0.8$, using data from the the NRAO VLA Sky Survey \citep[NVSS;][]{Condon98}. We assumed $\alpha=0.8$ and calculated $\theta_\phi$ for the subset of simulated data we used. Using all the sources down to the limit of $10\,$nJy we found $\theta_\phi=1.6\times10^{-5}\,$deg or $0.06\,$arcsec. The sources are certainly clustered on the scale of our beam ($\simeq 8\,$arcsec), but very weakly, because $\theta_\circ$ (the angular scale for non-linear clustering) is so small compared with our synthesized beam size. Our \textit{P(D)} calculation does not take into account any clustering correction. It also does not account for source sizes, but assumes that all the sources are unresolved. While in the case of both the VLA data and the simulated data, many of the sources are smaller than the beam, we know that this is not the case for all of them. In the simulated data there are roughly 700,000 sources, with a mean major axis size of $1.4 \,$arcsec and mean minor axis of $0.8 \,$arcsec, giving a mean source solid angle of $\Omega \simeq 1.27 \,$arcsec$^2$ (before convolution with the beam). This is much smaller than our beam solid angle of $72.3\,$arcsec$^2$.  

To test what kind of effect source sizes and clustering have on the model fitting, two other images were made. The first kept the source position information, so that any clustering would be preserved, but all source size information was neglected. Every source, single and multi-component, was set to a single delta function with flux density equal to the total source flux density, and then this was convolved with the beam. The second image also had all the sources as delta functions, but in this case the positions were randomised as well, so that the sources were unclustered. The MCMC fitting was rerun on histograms from these two simulated images with all other factors being the same. The results from fitting each of the three images are compared in the bottom panel of Fig.~\ref{fig:sim_counts}.  

The amplitudes and error regions of the three brighter $S$ nodes are the same in each case. The only differences are for the faintest nodes, which are more difficult to constrain. Comparing the full image with the case where no size information is present, we see that the full image case is higher; as anticipated given that with the larger source sizes one might expect more blending, more bright pixels, and a slightly wider histogram. When comparing the case with randomised positions and unresolved sources, again the results are the same down to about the noise level, although fainter than this does give lower values. This again is expected, due to the lack of both source sizes and clustering. Clustering within the beam will tend to boost the pixel values after beam convolution, producing a slight widening of the distribution \citep[see][]{Takeuchi04}. We would expect these fainter nodes to be of lower amplitude without clustering, as seen. When not accounting for source sizes or clustering, the largest fractional change in node amplitude from the full image is $2.3\,$per cent at the first node, $2.2\,$per cent for the second node, with the others all $1\,$per cent or less; all of the values lie within the $68\,$per cent confidence limits of the full simulation. These results make us confident that neglecting the effects of clustering and source size when fitting our real data results in no significant bias. 

Regarding the issue of source sizes, it is important to note that \textit{P(D)} counts are much more robust than comparably deep individual source counts. This is because \textit{P(D)} counts use a much bigger beam. For example, our $8\,$arcsec \textit{P(D)} beam corresponds to about one source per beam. Individual sources can be counted reliably only if there are at least 25 beams per source. This means the beam width for individual counts can not be much bigger than $8./\sqrt{25}\sim 1.6\,$arcsec, which is quite close to the mean source size in the SKADS simulation and would require large corrections for partial resolution of the sources.

\section{Results}
\label{sec:results}
\subsection{Estimated number counts}
\label{sec:numbers}
For all the models investigated here we report the means from the marginalised parameter likelihood distributions for the variable parameters and any derived parameters. This is done both for fitting just the first noise zone and for fitting all three zones simultaneously. The limits listed are $68\,$per cent (upper and lower) confidence limits for the marginalised means, except for the first node which is only an upper limit. We can also compare these results with the single power-law best-fit from \citetalias{Condon12} and with a compilation of known source counts from \citet{dezotti09}. The confusion noise is measured from the noiseless \textit{P(D)} distribution (eq.~(\ref{eq:pofd1}) with the noise term set to zero). Calculating the standard deviation is not an accurate way of finding $\sigma_{\rm c}$, since it is such a skewed distribution. Instead we found the median and $D_1$ and $D_2$ such that
\begin{equation}
 \sum_{D_1}^{\rm median} P(D) =\sum_{\rm median}^{D_2} P(D) =0.34
 \label{eq:confuse}
 \end{equation}
when normalised such that the sum of the $P(D)=1$, since in the Gaussian case $68\,$per cent of the area is between $\pm 1\sigma$. Then we took $\sigma_{\rm c}=(D_2-D_1)/2$. The confusion noise values for the different models are listed in Tables~\ref{tab:modpl} and \ref{tab:params}. The value estimated from the single power-law fit in \citetalias{Condon12} is $1.2 \, \mu$Jy beam$^{-1}$, in the middle of our range of $1.05 \leq \sigma_{\rm c}^{*}\leq 1.37 \, \mu$Jy beam$^{-1}$.

\begin{figure}
\includegraphics[scale=.375,natwidth=9in,natheight=9in]{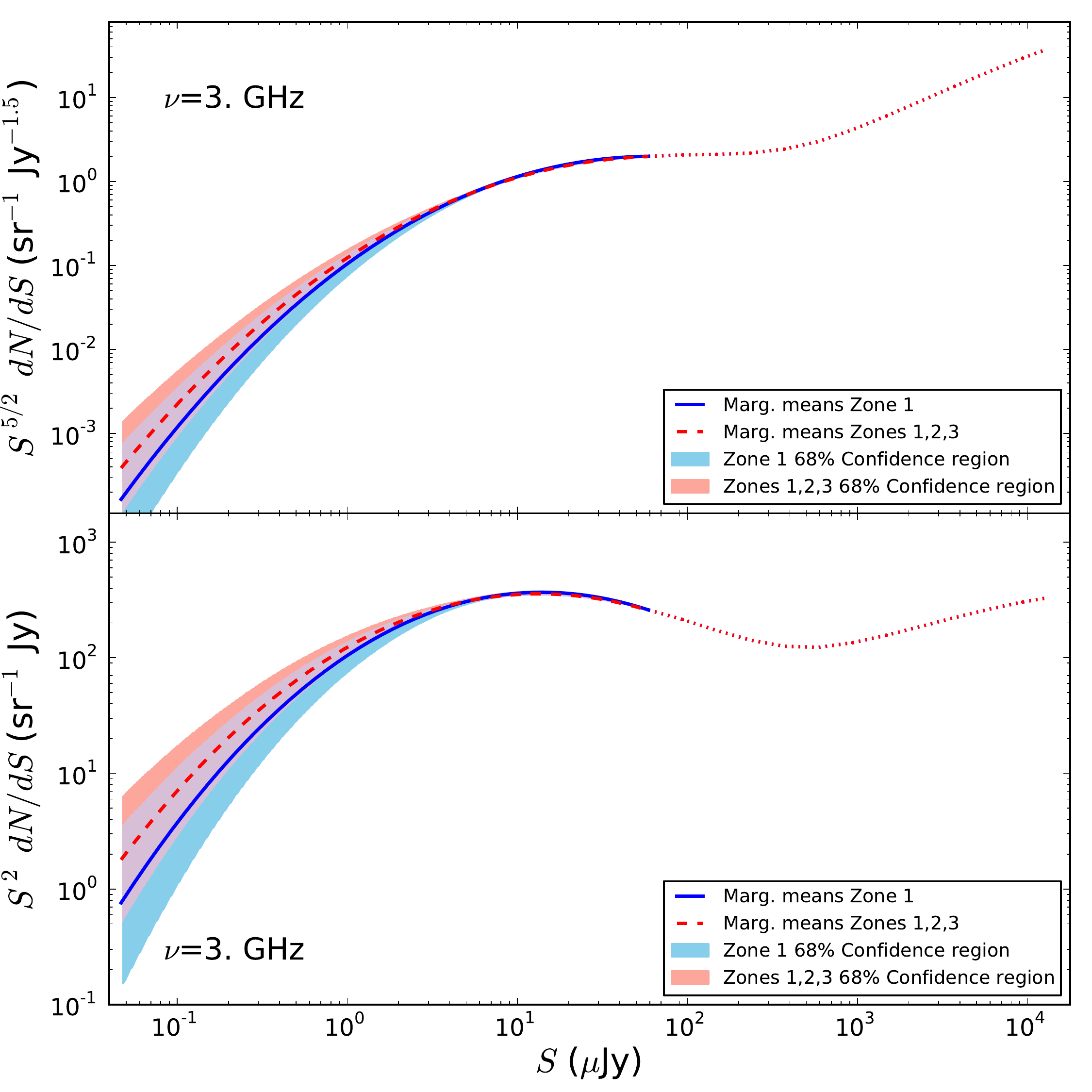}
\caption{Source count at $3 \,$GHz from MCMC fitting of the modified power-law model from eq.~(\ref{eq:powerl}). Lines are from the marginalised means of the parameters of eq.~(\ref{eq:powerl}) (red dashed is from all three zones, i.e. out to $10\,$arcmin, while blue solid is from zone 1, i.e. $5\,$arcmin). The dotted lines is where the model was fixed to the values of the Condon (1984) model. The shaded areas are $68\,$per cent confidence regions. The top panel uses the Euclidean normalisation, while the bottom panel has the $S^2$ normalisation. }
\label{fig:comparempld}
\end{figure}

\begin{figure}
\includegraphics[scale=0.375,natwidth=9in,natheight=9in]{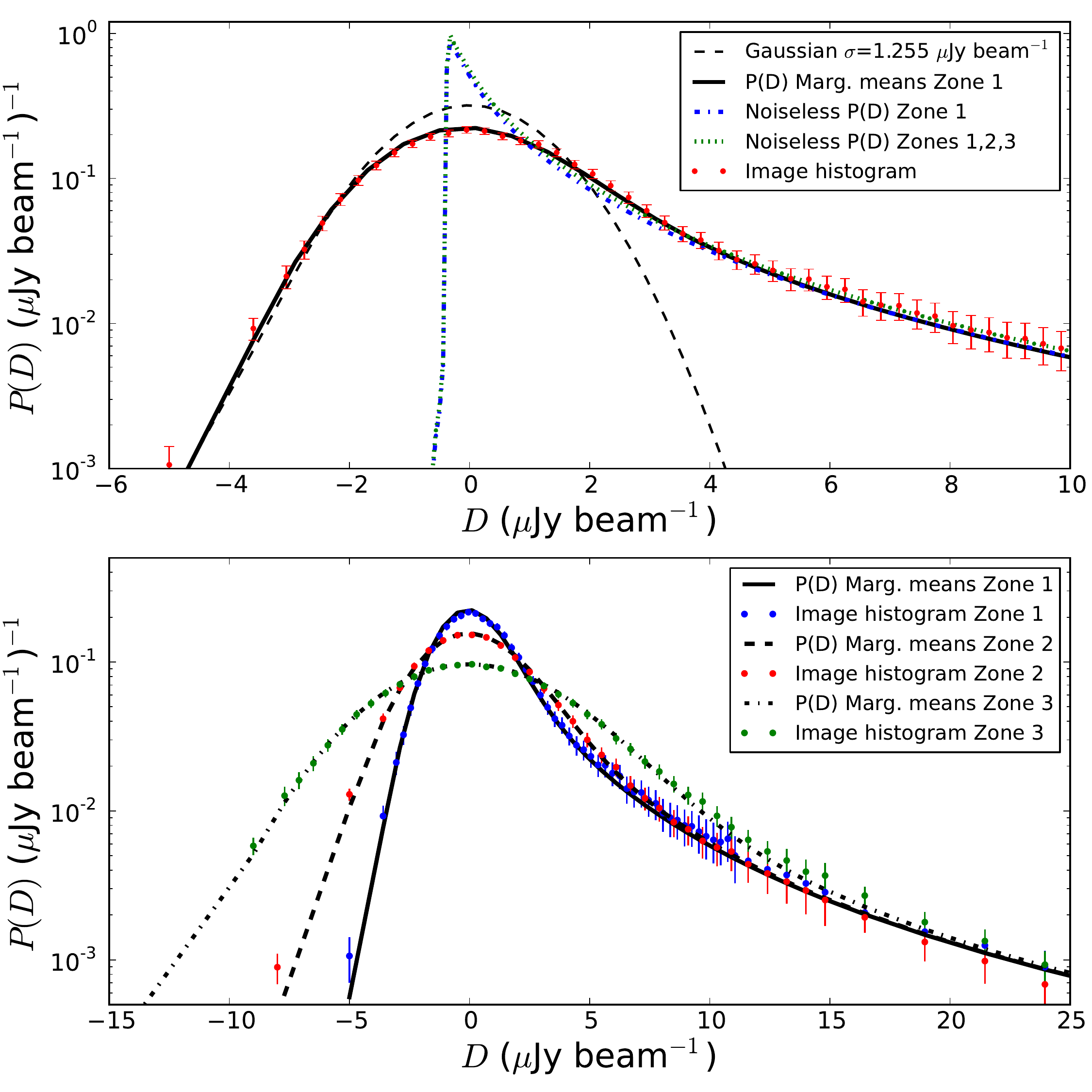}
\caption{ Comparison of $3 \,$GHz pixel histograms (red dots) with the marginalised means model \textit{P(D)} for zone 1 only (top panel) and models for all three zones (bottom panel) using a modified power-law input model. The dashed line is Gaussian noise of $\sigma = 1.255 \, \mu$Jy beam$^{-1}$. The noiseless \textit{P(D)} for each model is shown by the blue dot-dashed line for the one zone fit and the green dotted line for the three zone fit. }
\label{fig:pdflogmpl}
\end{figure}

The MCMC fitting was first run with the modified power-law model. The results from these runs are listed in Table~\ref{tab:modpl}, with the fits scaled to $1.4\,$GHz plotted in Fig.~\ref{fig:comparempld}. The data and model \textit{P(D)} distributions can be seen in Fig.~\ref{fig:pdflogmpl},  along with the noise distributions and model noiseless \textit{P(D)} distributions. Above about $3 \,\mu$Jy all the fits are consistent. Below this the results from fitting the three noise zones simultaneously fall off faster than the fits from the first noise zone alone. 

\begin{table}
  \caption{Marginalised fits for the modified power law in eq.~(\ref{eq:powerl}) at $3\,$GHz. The quoted uncertainties are $68\,$per cent confidence intervals. For the combined fit we treat each zone separately, and hence the number of degrees of freedom is approximately 3 times higher.}
  \begin{tabular}{lll}
\hline\hline
Noise zones&1&1, 2, 3\\
\hline
Parameter&Marginalised means&Marginalised means\\
$\alpha$ & $-4.5^{+1.3}_{-1.3}$& $-4.7^{+1.2}_{-1.2}$\\
$\beta$& $-0.17^{+0.25}_{-0.25}$ &$-0.16^{+0.25}_{-0.25}$ \\
$\gamma$ &$0.012^{+0.017}_{-0.017}$ &$0.016^{+0.016}_{-0.016}$ \\
$\log_{10}(\kappa)$ &$-4.34^{+1.3}_{-1.3}$&$-5.01^{+1.2}_{-1.1}$\\
\hline
 $\sigma_{\rm c} \,$($\mu$Jy beam$^{-1}$)   &$1.122^{+0.009}_{-0.009}$  &$1.068^{+0.008}_{-0.008}$ \\
$ \chi^2$  & 87.3 &160.3 \\
$N_{\rm dof}$  &59&149\\
\hline
\end{tabular}
\label{tab:modpl}
\end{table}

\begin{figure}
\includegraphics[scale=0.375,natwidth=9in,natheight=9in]{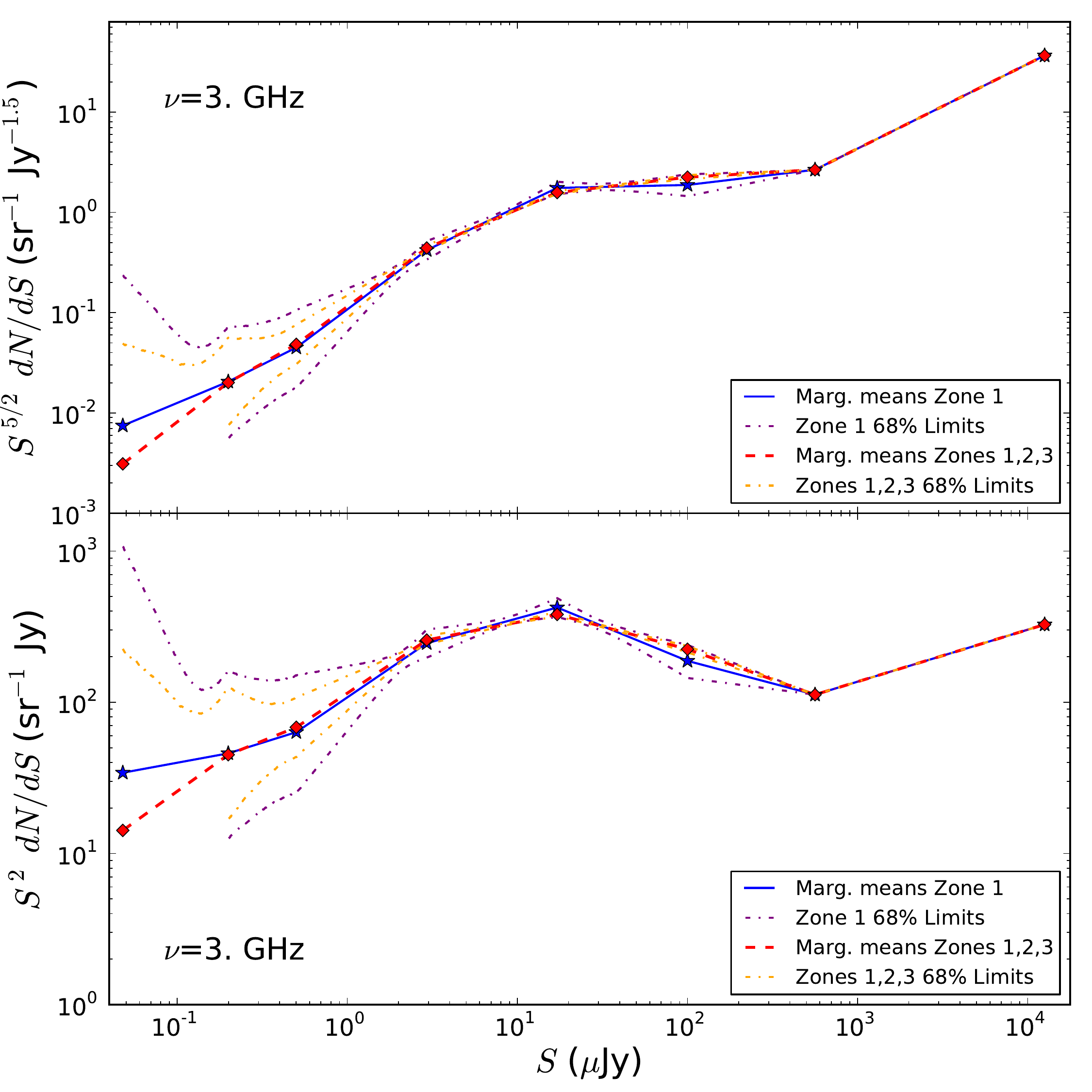}
\caption{ Source count at $3 \,$GHz from MCMC fitting of the node-based model using six free nodes and two fixed nodes. Points and corresponding lines are the node marginalised means, with the red dashed line being from all three noise zones (out to $10\,$arcmin), while the blue solid line is from one zone ($5\,$arcmin). The dot-dashed lines are $68\,$per cent confidence regions (purple for Zone 1, orange for all three zones). The top panel uses the Euclidean normalisation, while the bottom panel has the $S^2$ normalisation indicative of contribution to the background temperature.}
\label{fig:comparesig}
\end{figure}

\begin{figure}
\includegraphics[scale=0.375,natwidth=9in,natheight=9in]{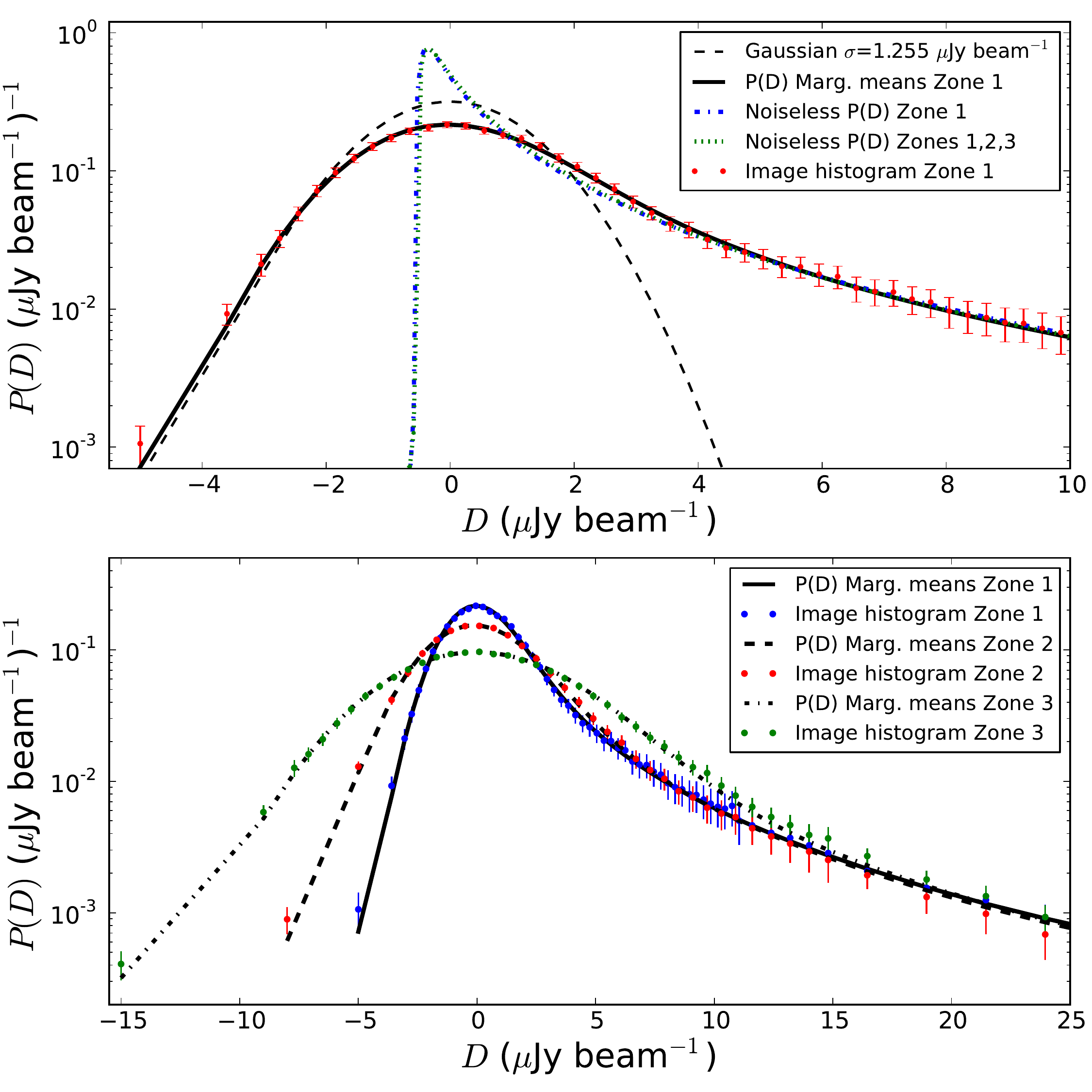}
\caption{Comparison of $3 \,$GHz pixel histograms (red dots) with the marginalised means model \textit{P(D)} for Zone 1 only (top panel) and models for all three zones (bottom panel) for the node-based input model. The dashed line is Gaussian noise of $\sigma = 1.255 \, \mu$Jy beam$^{-1}$. The noiseless \textit{P(D)} for each model is shown by the blue dot-dashed line for the one zone fit and the green dotted line for the three zone fit.}
\label{fig:pdflog}
\end{figure}

The results for the node-based model are listed in Table~\ref{tab:params}. The slopes and normalisation constants for the interpolated power laws between the nodes, of the form ${dN/dS} = kS^{\gamma}$, are listed in Table~\ref{tab:slope}. The source counts from these models are plotted in Fig.~\ref{fig:comparesig} and the \textit{P(D)} distributions are shown in Fig.~\ref{fig:pdflog}. The $\chi^2$ values are lower than in the modified power-law model, though the $\chi^2$ values for all four model fits are reasonably consistent with $N_{\rm dof}$ the number of bins minus the number of fit parameters. The models are consistent with each other, except for $S\le 1\, \mu$Jy, where the node-based model falls off more slowly. The modified power law has the advantage of being a single continuous function, as well as having less fit parameters. However, the node-based model allows for a larger range of possibilities than the modified power law and is much more sensitive to the count below the noise level, as it is able to fit that region with little to no effect on the brighter values. With this model, the count for the one-zone case is above those from the three-zone case in the faint region, although the marginalised means are almost identical. 

\begin{table}
  \caption{Marginalised mean amplitudes for the six fit nodes and two fixed nodes at $3 \,$GHz, given separately for the deepest noise zone and for all three noise zones fit simultaneously. The brightest two nodes were fixed to values estimated from known counts at $1.4\,$GHz and scaled to $3 \,$GHz using $\langle\alpha\rangle = -0.7$.  }
  \begin{tabular}{lll}
\hline 
\hline
 Noise Zones&1&1, 2, 3\\
 \hline
Node&Marginalised means &Marginalised means\\
$\mu$Jy&$\log_{10}$[sr$^{-1}$ Jy$^{-1}$]&$\log_{10}$[sr$^{-1}$ Jy$^{-1}$]\\
\hline
0.05& $16.17^{+1.69}$&$15.79^{+1.20}$\\
0.20 &  $ 15.06^{+0.56}_{-0.56} $ &  $15.05^{+0.45}_{-0.43} $\\
0.50 & $14.43^{+0.38}_{-0.40}$  & $14.43^{+0.20}_{-0.20}$\\
2.93 & $13.45^{+0.09}_{-0.09} $ & $13.48^{+0.03}_{-0.03}$\\
17.2 & $12.16^{+0.06}_{-0.06}$  & $12.11^{+0.02}_{-0.02}$\\
100 & $10.27^{+0.11}_{-0.11} $ & $10.35^{+0.02}_{-0.02}$\\
572 & 8.55  & 8.55\\
12600& 6.32  & 6.32\\
\hline
 $\sigma_{\rm c} \,$($\mu$Jy beam$^{-1}$)  &$1.283^{+0.006}_{-0.007}$  &$1.266^{+0.003}_{-0.003}$ \\
$ \chi^2$&54.8&153.05\\
$N_{\rm dof}$ &59&149\\
\hline
\end{tabular}
\label{tab:params}
\end{table}

\begin{table}
  \caption{Slopes and normalisation constants for the interpolated power laws between the nodes, of the form ${dN \over dS} = kS^{\gamma}$ at $3 \,$GHz.  }
  \begin{tabular}{lcccc}
\hline 
\hline
 Noise Zones &\multicolumn{2}{c}{1}&\multicolumn{2}{c}{1, 2, 3}\\
 \hline
Between &\multicolumn{2}{c}{Marginal fit}&\multicolumn{2}{c}{Marginal fit}\\
Nodes ($\mu$Jy)  &  $\gamma$ & $\log_{10}k_{\rm 3 GHz}$ &$\gamma$& $\log_{10}k_{\rm3 GHz}$ \\
\hline
$0.05{-}0.20 $&$-1.79$&$3.05$ &$-1.19 $&$7.06$\\
$0.20{-}0.50$& $-1.65$ &$4.01$& $-1.55$   &$4.69$  \\
$0.50{-}2.90 $& $-1.23$ &$6.63 $ &$-1.25$&$6.57$\\
$2.93{-}17.2$& $-1.69$ &$4.09$ &$-1.78$&$3.63$\\
$17.2{-}100$&$-2.46$&$0.43$ &$-2.30$&$1.15$\\
$100{-}560$  &$-2.29$&$1.08$ & $-2.40$&$0.75$\\
$572{-}12600$  &$-1.66$ &$3.16$ & $-1.66$&$3.16$\\

\hline
\end{tabular}
\label{tab:slope}
\end{table}

\subsection{Parameter degeneracies}
\label{sec:degen}
The values of the parameters are highly correlated, particularly between adjacent nodes where the correlation is negative. This means that the errors on the number count parameters will also be correlated, giving non-Gaussian shapes to some of the joint likelihoods of the two parameter distributions. Sources at a given flux density contribute to many different \textit{P(D)} pixel values when convolved with the beam. This means that some sources could be effectively moved from one flux density bin to another, still retaining the same shape for the resulting histogram. This is illustrated in the confidence regions plotted with the source counts (see Fig.~\ref{fig:compare}). Instead of being straight power laws from one parameter's upper limit to the next, the confidence regions tends to ``bow" inwards between the two nodes; as one node amplitude is raised the amplitude of the neighbours must decrease. This degeneracy is strongest for the fainter flux densities, as they are not only degenerate with neighbouring nodes, but also with the instrumental noise. 

\begin{figure*}
\includegraphics[width=6.5in,height=6.5in,natwidth=12in,natheight=12in]{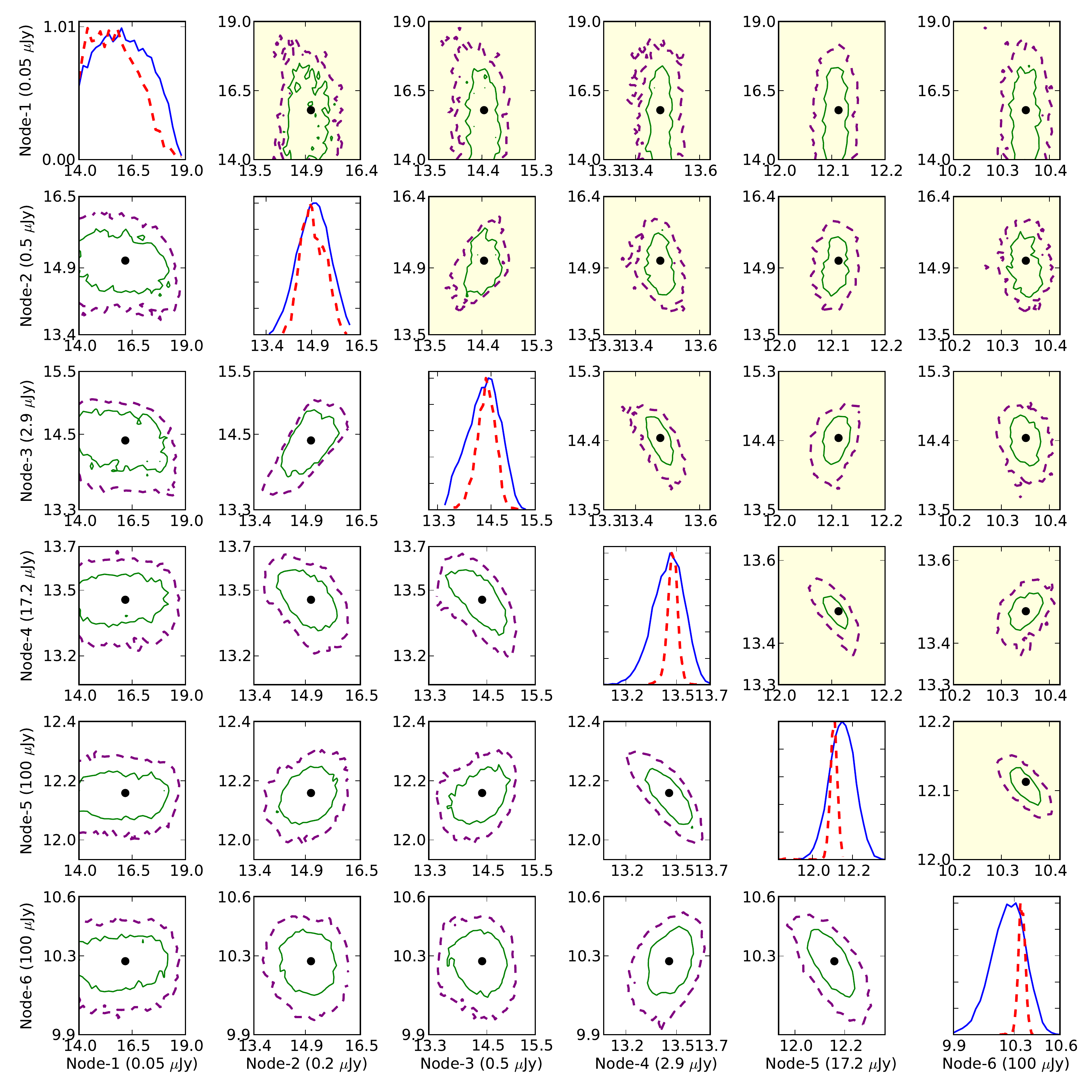}
\caption{One  and two dimensional likelihood distributions for the six fit nodes. The upper triangle 2D plots (yellow background) are for the three noise zone fits and the lower triangle 2D plots are the one noise zone fits. The 1D plots show the marginalised likelihood distributions for those nodes with three noise zones (red dashed line) and one noise zone (blue solid line). For the 2D plots the contours are $68$ (green solid) and $95\,$per cent (purple dashed) confidence limits. The black dots show the positions of the marginalised means. Parameter units for each plot are $\log_{10}$[sr$^{-1}$Jy$^{-1}$].}
\label{fig:tri}
\end{figure*}

The Pearson correlation matrix for the two cases is listed in Table~\ref{tab:correl}, and the 2D likelihood distributions are shown in Fig.~\ref{fig:tri}. The degeneracy means that adding more nodes in the fainter regions does not improve the fit. We would require lower instrumental noise, as well as increased resolution, to benefit from extra nodes. 

\begin{table}
\small
  \caption{Correlation matrix for parameters. Coefficients are computed for fitting all three zones (upper triangle) and just zone one (lower triangle), following the definition $C_{ij}=\sum_r p_ip_j / \sqrt{\sum_r p_i^2 \sum_r p_j^2}$, where $p_i$ and $p_j$ are parameter numbers $i$ and $j$, and $r$ is the realization number.}
 \begin{tabular}{cr@{.}lr@{.}lr@{.}lr@{.}lr@{.}lr@{.}l}\\
\hline\hline
Node ($\mu$Jy) & \multicolumn{2}{c}{$0.05$}&\multicolumn{2}{c}{$0.20$}&\multicolumn{2}{c}{$0.50$}&\multicolumn{2}{c}{$2.90$}&\multicolumn{2}{c}{$17.2$}&\multicolumn{2}{c}{$100$}\\
\hline
0.05 & 1&00 & $-0$&$16$ & $-0$&$17$ &$0$&$02$ & $0$&$03$&$0$&$01$ \\
0.20 &$-0$&$23$&1&00&$0$&$35$&$-0$&$28$&$0$&$15$&$-0$&$11$\\
0.50 &$-0$&$26$ &$0$&$66$&$1$&$00$ & $-0$&$64$ & $0$&$33$ & $-0$&$19$ \\
2.93 &$0$&$01$& $-0$&$44$&$-0$&$61$& $1$&$00$ & $-0$&$78$& $0$&$44$\\
17.2 &$0$&$03$& $0$&$25$ &$0$&$31 $&$-0$&$79$& $1$&$00$& $-0$&$72$\\
100 &$0$&$01$ &$ -0$&$11$ &$- 0$&$14$ & $0$&$37$ &$-0$&$68$& $1$&$00$\\
\hline
\end{tabular}
\label{tab:correl}
\end{table}
\normalsize

\subsection{Background temperature}
\label{sec:temp}

We can obtain an estimate of the contribution from discrete sources to the radio background temperature from our new source count. The source count and the sky temperature at a frequency $\nu$ are related by the Rayleigh-Jeans approximation,
\begin{equation}
\int_{S_{\rm min}}^\infty S \frac{dN}{dS} dS=  {T_{\rm b} c^2 \over 2 k_{\rm B}\nu^2 }. 
\label{eq:integrate}
\end{equation}
In the above equation $k_{\rm B}$ is the Boltzmann constant, and $T_{\rm b}$ is the sky temperature from all the sources brighter than $S_{\rm min}$.  Equation~(\ref{eq:integrate}) is also equivalent to
\begin{displaymath}
\int_{S_{\rm min}}^\infty S^2 \frac{dN}{dS} d[\ln(S)]=  {T_{\rm b} c^2 \over 2 k_{\rm B}\nu^2 }. 
\end{displaymath}
It is for this reason that in Figs.~\ref{fig:comparempld} and \ref{fig:comparesig} the bottom panels show the source count weighted not by the Euclidean $S^{5/2}$ but by $S^{2}$. This alternate weighting of $S^{2}dN/dS$ is proportional to the source count contribution to the background temperature per decade of flux density. With such a plot the source count must fall off at both ends to avoid violating Olbers paradox; for this reason it is also the case that the brighter side must turn over at flux densities brighter than we have plotted. 

\begin{figure}
\includegraphics[scale=0.375,natwidth=9in,natheight=9in]{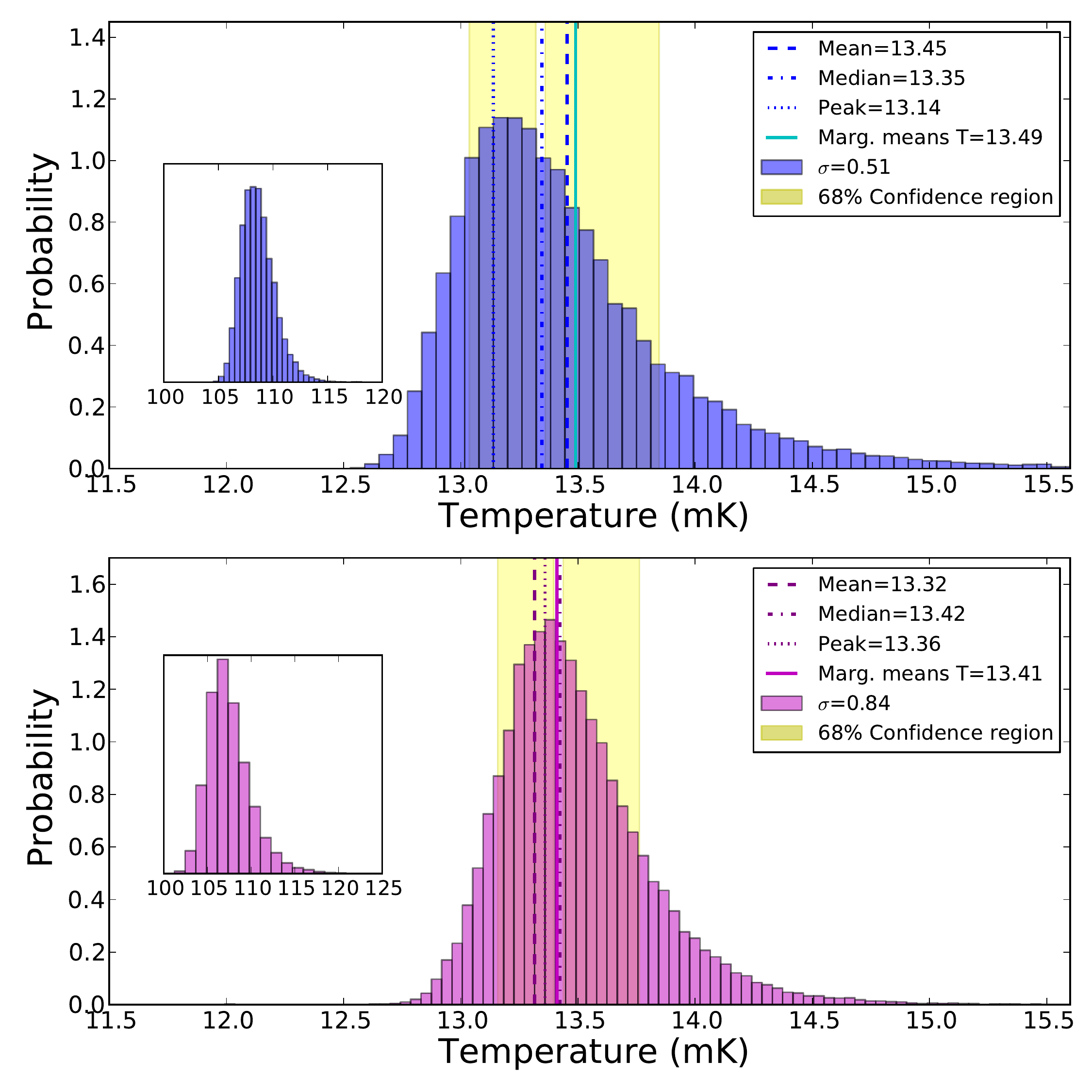}
\caption{Normalised histogram of radio background temperatures at $3\,$GHz from integrating each step in the MCMC according to eq.~(\ref{eq:integrate}) using the {\it modified power-law model}. The top panel comes from just fitting noise Zone 1, while the bottom panel is from fitting for all three noise zones. Insets are temperature histograms at $1.4\,$GHz made by scaling the $3\,$GHz chains with $\langle\alpha\rangle = -0.7$.   }
\label{fig:temphist3mpl}
\end{figure}

\begin{figure}
\includegraphics[scale=0.375,natwidth=9in,natheight=9in]{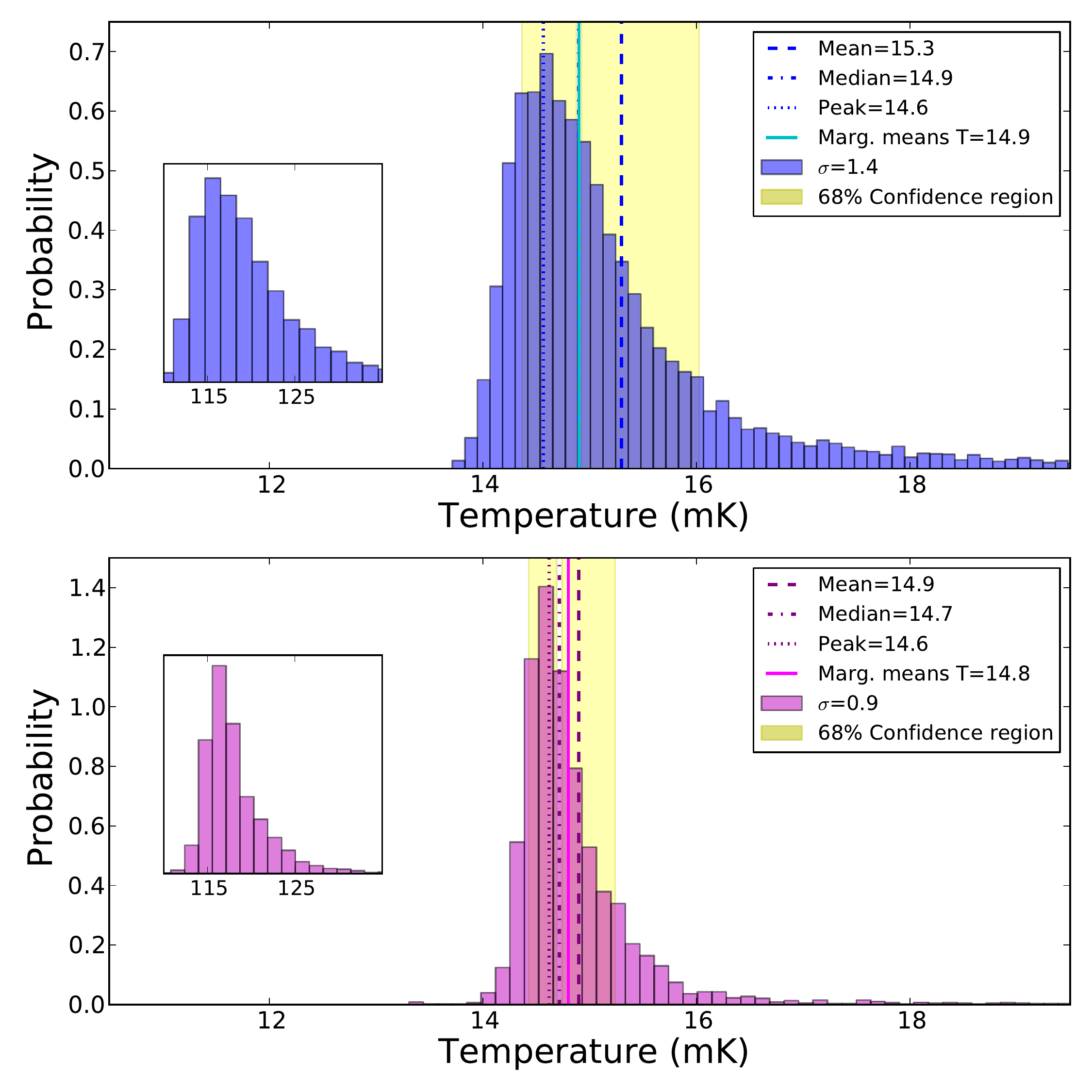}
\caption{Normalised histogram of radio background temperatures at $3\,$GHz from integrating each step in the MCMC according to eq.~(\ref{eq:integrate}) using the {\it node-based model}. The top panel comes from just fitting noise Zone 1, while the bottom panel is from fitting for all three noise zones. Insets are temperature histograms at $1.4\,$GHz made by scaling the $3\,$GHz chains with $\langle\alpha\rangle = -0.7$.   }
\label{fig:temphist3}
\end{figure}

Using eq.~(\ref{eq:integrate}) we are able to obtain estimates for the discrete-source contribution to the background temperature from our results. Integrating the MCMC output at each step in the chains allows us to look at the distribution of temperatures. Fig.~\ref{fig:temphist3mpl} shows the histograms obtained from the modified power-law fitting for both noise zone cases at $3 \,$GHz, as well as scaled to $1.4 \,$GHz; the same is shown in Fig.~\ref{fig:temphist3} for the node-based model. For the modified power-law fits we integrated over the flux density range $0.05 \leq S\,(\mu$Jy$)\leq 60$ and used the values from the Condon (1984) model for $60\,{< }\,S\, (\mu{\rm Jy})\,{<}\,10^{9}$.  For the node-based model the fit results were used in the range $0.05\,{<}\,S\,(\mu {\rm Jy})\,{<}\,1.26\times10^{4}$ and the Condon (1984) model for $1.26\times10^{4}\,{<}\,S\,(\mu {\rm Jy}) \,{<}\,10^{9}$.

The outputs obtained from the MCMC fitting allow us to compute $68\,$per cent confidence intervals for each distribution, as well as the means, medians, peaks, and values from the source counts from the marginalised means from each parameter. These values are listed in Table~\ref{tab:temp}. The values from the different models and noise settings are all consistent. These yield a background temperature of around $14.5 \,$mK at $3 \,$GHz, corresponding to $115 \,$mK at $1.4 \,$GHz. The distributions from the node-based models tend toward higher values and have more elongated tails. Because of this skewness, the $68\,$per cent confidence limits for these two distributions are computed from the median instead of the mean. This skewness is simply due to the fact that this model allows for more possible values in the faintest region, letting the faintest node rise to higher amplitudes, thus affecting the integrated temperature. 

\begin{table*}
\begin{minipage}{110mm}
  \caption{Radio background temperatures from integration of the source counts using eq.~(\ref{eq:integrate}).}
  \begin{tabular}{llllllllll}
\hline\hline
 Model &\multicolumn{4}{c}{Node model}&\multicolumn{4}{c}{Modified power-law model}\\
 \hline
Frequency (GHz)&3.0&3.0&1.4&1.4&3.0&3.0&1.4&1.4\\
Zones & 1&1, 2, 3&1&1, 2, 3& 1&1, 2, 3&1&1, 2, 3\\
\hline
Peak (mK)& 14.6&14.6&115.8&116.3& 13.1&13.4&104.7&106.1\\
Median (mK)&14.9&14.7&118.7&117.3& 13.3&13.4&106.4&106.9\\
68$\%$ Lower Limit (mK)&14.4&14.4&115.5&115.5& 13.1&13.2&103.9&104.9\\
68$\%$ Upper Limit (mK)&16.4&15.3&127.7&121.5& 13.9&13.8&110.4&109.9\\
marginalised Fit (mK)&14.9&14.8&109.2&111.7& 13.5&13.4&107.7&106.8\\
\hline
\end{tabular}
\label{tab:temp}
\end{minipage}
\end{table*}

\section{Discussion}
\label{sec:dis}
\subsection{Systematics}
\label{sec:sys}
\subsubsection{Image artefacts}
\label{sec:sysnoise}
\begin{figure}
\includegraphics[scale=0.375,natwidth=9in,natheight=9in]{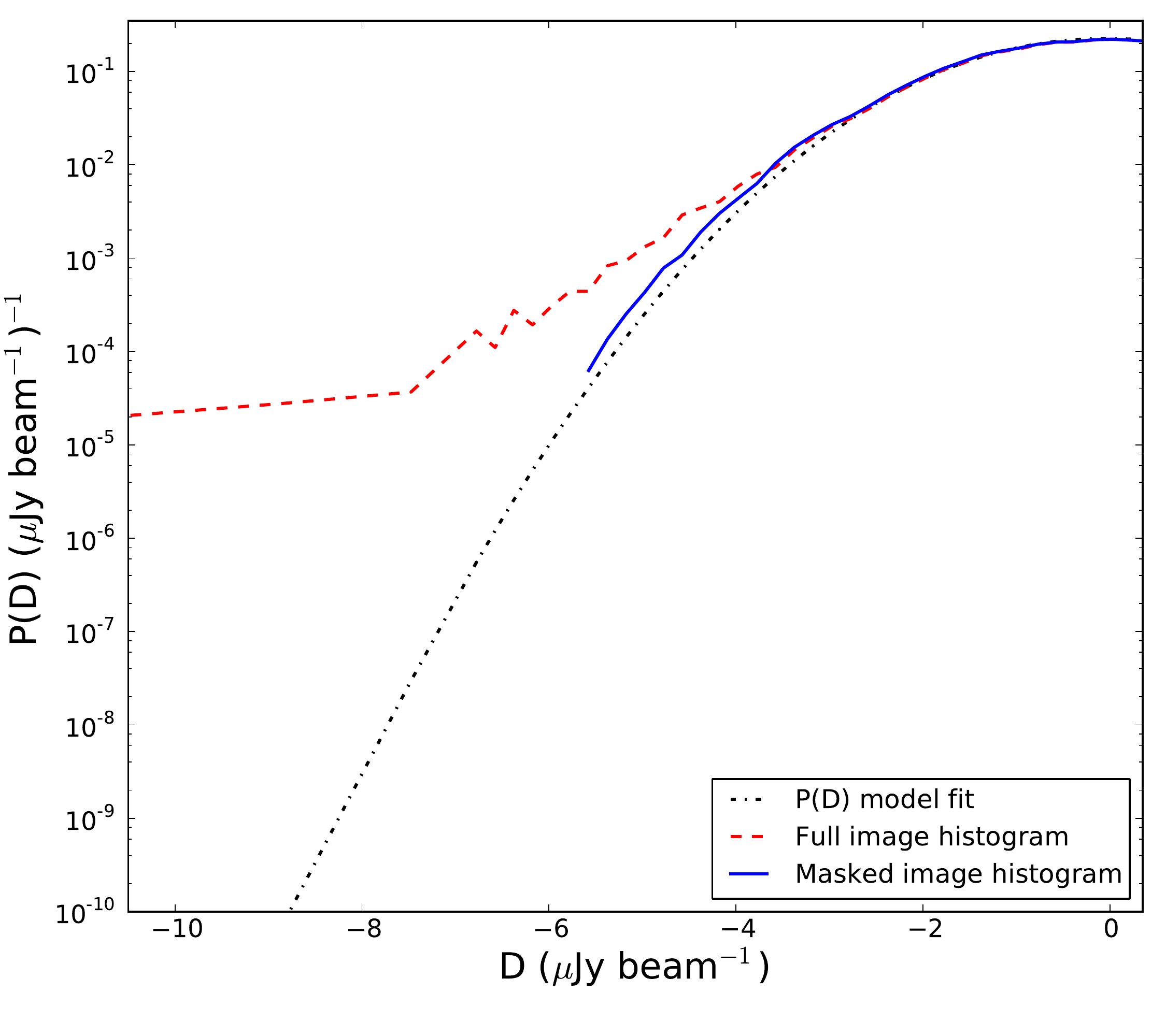}
\caption{Negative flux-density region of the \textit{P(D)} distribution. The red dashed line is from the full histogram of the data. The black dot-dashed line is from the marginalised mean node parameters from the MCMC fitting. The blue solid line shows the image histogram after masking.}
\label{fig:negflux}
\end{figure}
This \textit{P(D)} fitting technique assumes that the instrumental noise is Gaussian and well characterised. In practice our image noise is very nearly Gaussian, with the highest contamination from dirty beam sidelobes being only about $0.1 \, \mu$Jy beam$^{-1}$. There is, however, another effect that contributes to the shape of the histogram: it appears to have a tail of excess negative flux density pixels and thus does not drop off in a purely Gaussian way on the negative side. When looking at the image it is clear that the pixels responsible (with values $>5\sigma$, even considering the primary beam correction) are all clustered around the brightest two sources, with almost all of them around the brightest source in the image which is about $7\,$mJy, right at the edge of the $5\,$arcmin ring. 

It is clear that this is an artefact caused by the imaging and cleaning process, or by asymmetry in the antenna pattern. The VLA antennas use alt-az mounts which cause the antenna pattern to rotate on the sky with parallactic angle. The support legs for the secondary introduce asymmetries in the antenna pattern, which, when combined with the rotation with parallactic angle, cause sources away from the pointing centre to appear variable. This in turn causes areas of negative excess pixels around brighter sources. The effect gets stronger with distance from the field centre, with increasing frequency, and source brightness. When examining the 16 sub-band images it does seem that this effect increases in strength with frequency. However, it is difficult to say if this is truly the cause, as beyond $5\,$arcmin there are very few bright sources. To be able to remove this effect we would need detailed measurements of the antenna beam pattern at S-band. Such measurements have not yet been made; we hope that future imaging will be able to correct for artefacts of this type. 

With the current data the presence of this negative tail seriously affects the MCMC fitting. Since there is a large deviation from the predicted \textit{P(D)} calculation with Gaussian noise, attempting to fit the entire histogram gives too broad a distribution, and too low a peak. The fitting procedure inflates the faintest (and possibly second faintest) node to higher amplitudes to achieve this. Without being able to correct or model the antenna pattern we simply masked out the negative pixels around the $7\,$mJy source and a smaller region near a second ($1\,$mJy) source. This decreased the total number of pixels used by about $0.2\,$per cent. We also masked the negative pixels in the second noise zone around this source, decreasing its number of pixels by $0.07\,$per cent.

The negative side of the image \textit{P(D)} can be seen in Fig.~\ref{fig:negflux}. Red points indicate the image values, while the black line is the \textit{P(D)} model using the node values from the first noise zone fitting. The blue line shows the image values after masking out the negative regions. Tests run on the masked and unmasked versions clearly show that the masking has no effect on any nodes other than the first two, which become artificially inflated in the unmasked fitting. Thus we feel justified in performing the masking. All of the results presented in Section~\ref{sec:results} were fit using the masked images.

\subsubsection{Weighting}
\label{sec:weight}
New data reduction and imaging challenges arise from the $2\,$GHz bandwidth of the VLA at S band. Across this bandwidth there are substantial changes in the synthesised beam size and the primary beam, as well as source flux-density changes due to the spectral dependence. In our particular case each sub-band was imaged independently (although cleaned simultaneously), with weighting and taper factors applied during cleaning to force the synthesized beams to be the same size. In the narrow-band case changes due to the frequency bandwidth are usually small and thus weighting to produce an image at the centre frequency of the band does not usually need to include any spectral dependence. With wide-band data this type of weighting scheme would maximize signal to noise only for sources with $\langle\alpha\rangle = 0$. Instead one could perform a weighted fit of the spectral dependence in each pixel of the 16 sub-band images, correct for the primary beam spectral dependence at the distance of each pixel from the centre, and use that value to calculate the flux density at the centre frequency. However, this requires having enough signal-to-noise in each pixel to obtain an accurate fit. The weighting scheme we used was
\begin{equation}
 W_{i}(\rho, \nu_{i}) \propto \biggl[{\nu_{\rm c}^{\langle\alpha\rangle}
 \over \sigma_{\rm n\it i} A(\rho,\nu_{\rm c})}\biggr]^2~,
\end{equation}
where $i$ labels the sub-bands, $\sigma_{\rm n\it i}$ is the noise in each sub-band image, and $ A(\rho,\nu_{\rm c})$ is the primary beam value at pixel distance $\rho$ and sub-band frequency $\nu_i$.  Using these weights the $3$-GHz pixel values were given by
\begin{equation}
b_{\rm 3GHz}(\rho) = \sum_{i = 1}^{16}\, [b_i (\rho) W_i
  (\rho)]\,\bigg/\, \sum_{i=1}^{16} W_i(\rho),
\end{equation} 
with $b_i$ being the pixel brightness in the $i$th sub-band. This combination is designed to maximize the signal-to-noise ratio for sources with $\langle\alpha\rangle = -0.7$, the average spectral index for faint sources in this frequency range \citep[e.g.][]{Condon84b}. 

\begin{figure}
\includegraphics[scale=0.375,natwidth=9in,natheight=9in]{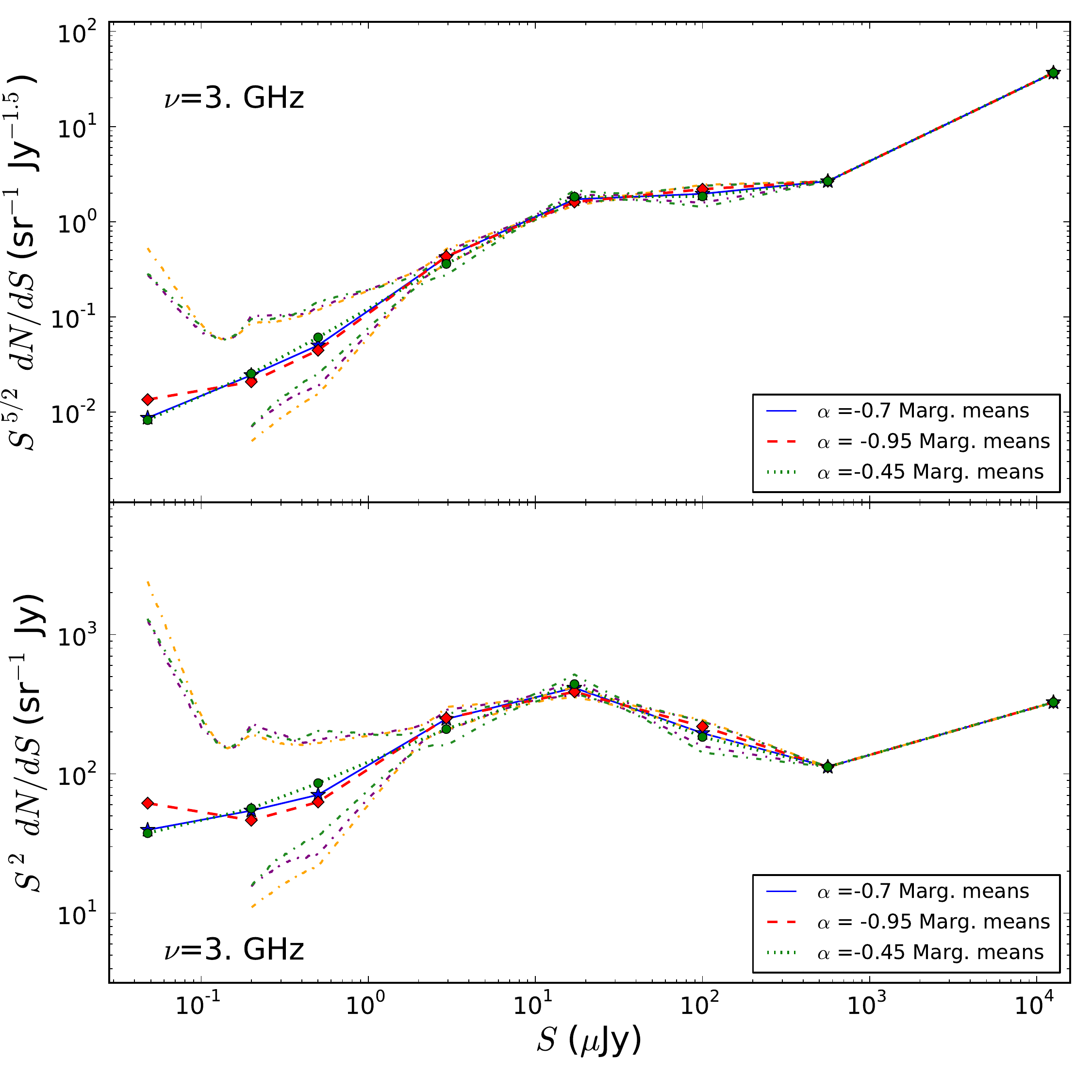}
\caption{Comparison of MCMC \textit{P(D)} fitting for wide-band images made with different spectral indices in the weighting scaled to $3 \,$GHz. The top panel is with the Euclidean normalisation, while the bottom panel is $S^2$ normalised. Lines  and points are the parameter marginalised means. The red dashed line is for $\langle\alpha\rangle = {-}0.95$,  the green dotted line is for $\langle\alpha\rangle = {-}0.45$, and the blue solid line is $\langle\alpha\rangle = {-}0.7$. The dot-dashed lines are the 68$\,$per cent confidence limits.}
\label{fig:alpha}
\end{figure}

However, it is possible that this choice of weighting scheme might have affected our \textit{P(D)} results. To test this we created two new wide-band images with different weightings applied. One image was made using $\langle\alpha\rangle = {-}0.45$ and one with $\langle\alpha\rangle = {-}0.95$. The MCMC fitting was rerun on both of these images, leaving the noise as a free parameter, since changing the weighting could have also affected the noise level. The marginalised mean values for the noise are $\sigma_{\rm n}^*=1.259 \, \mu$Jy beam$^{-1}$  for $\langle\alpha\rangle = {-}0.45$ and $\sigma_{\rm n}^*=1.245 \, \mu$Jy beam$^{-1}$  for $\langle\alpha\rangle = {-}0.95$. The results of the MCMC fitting can be seen in Fig.~\ref{fig:alpha}, compared with the $\langle\alpha\rangle = {-}0.7$ case with variable noise. There is very little difference in the fits. The largest fractional difference between the marginalised fits is still only $0.6\,$per cent for the third node between the $-0.7$ and the $-0.95$ cases. Therefore, it does not appear that the spectral dependence of the weighting has a significant effect on the output. 

\subsection{Comparison with other estimates}
\label{sec:compares}
\citetalias{Condon12} found that the best-fit slope for a single power-law in the $\mu$Jy region was $\gamma=-1.7$. It was noted that at the fainter end a shallower slope of $\gamma=-1.5$ or $-1.6$ might be better. Looking at the slopes between our fit nodes and the earlier result, we have three power-law sections that cover this region: between the second and third nodes corresponds to the region $0.2 \leq S\, (\mu {\rm Jy}) \leq 0.5$; between the third and fourth nodes is the region $0.5 \leq S\, (\mu {\rm Jy}) \leq 2.9$; and between the fourth and fifth nodes is the region $2.9 \leq S \, (\mu {\rm Jy}) \leq 17.2$. The slopes for these regions can be seen in Table~\ref{tab:slope}. For the faint part of the region our slopes range from $-1.23$ to $-1.65$, while for the brighter part they range from $-1.69$ to $-1.78$. The new results therefore agree well with a $-1.7$ slope for the brighter part of the $\mu$Jy region and do seem to suggest a shift to a shallower slope in the $\mu$Jy regime. Our $\chi^2$ value of $153.0$ for $ N_{\rm dof}=149$ over the full $10\,$arcmin is lower than those we obtain with a single-slope model of slope $-1.7$, where $\chi^2=249.1$ for $N_{\rm dof}=153$ (or a slope of $-1.6$, which gives $\chi^2=292.3$). By using the node-based model instead of the single power-law model the improvement in the fit yields $\Delta\chi^2=96.1$, which is a highly significant improvement for 149 degrees of freedom.

\begin{figure*}
\includegraphics[scale=0.45,natwidth=16in,natheight=8in]{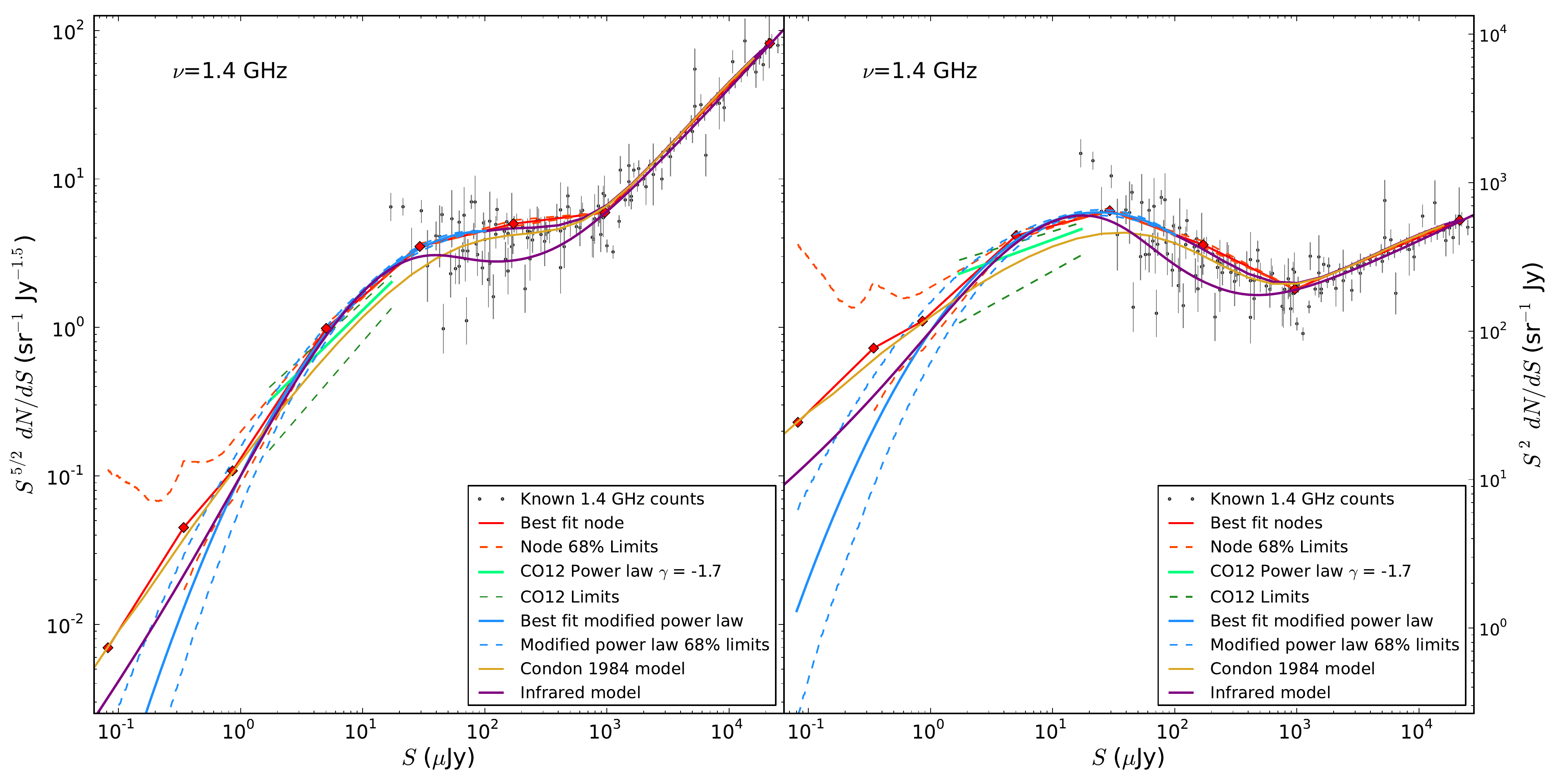}
\caption{Source counts at $1.4 \,$GHz of models and observed counts. Black points are known counts compiled from \citet{dezotti09}. The green line is a single power-law fit from \citetalias{Condon12}, with surrounding error box. The yellow line is the evolutionary model from \citet{Condon84b}. Red solid line and points are the best fit results from the MCMC fitting of the node-based model (three noise zones out to $10\,$arcmin). The dashed lines are $68\,$per cent confidence regions. The blue solid line is the best fit results from the MCMC fitting of the modified power-law model (three noise zones out to $10\,$arcmin). The results from this paper have been scaled to $1.4\,$GHz assuming a spectral index of $-0.7$. The purple solid line is the evolutionary model from \citet{Bethermin12b}.  The left panel uses the Euclidean normalisation, while the right panel has the $S^2$ normalisation.}
\label{fig:infrared}
\end{figure*}

\citet{Condon84b} used the local luminosity function to constrain the epoch-dependent spectral luminosity function of extragalactic radio sources, finding a simple model based on luminosity, redshift, and frequency that accurately predicted the source count at $1.4 \,$GHz at that time. This model shows two peaks in the $S^2dN/dS$ source count, one dominated by starburst-powered galaxies peaking at $50 \, \mu$Jy, and the other dominated by AGN-powered galaxies peaking near $0.1 \,$Jy. In the brighter flux density range, where there is a large amount of observational data, this model describes the source count well. We have plotted this Condon (1984) model against our fits for comparison in Fig.~\ref{fig:infrared}. At the brighter end of the count, $S \geq 90 \, \mu$Jy, there is good agreement between this model and all of our fits. In the region $S \leq 3 \, \mu$Jy the model is also within the uncertainties for all the node-based fits, and lines up quite closely with the marginalised mean fits. However, in the region $3 \leq S \, (\mu{\rm Jy}) \leq 90$ the Condon model is consistently below our fits, both from the node-based model and the modified power law. In this region the dominant component of the Condon model is star-forming galaxies. The discrepancy between the model and our source count results suggests the contribution from these galaxies is greater than previously thought. If the star-forming component from Condon's model is increased by roughly a factor of 2 it would match quite closely. It is clear that any successful models should not deviate too strongly from the Condon (1984) model, but may need a slightly different treatment of star-forming galaxies. 

We have also compared our results with the empirical model from \citet{Bethermin12b}. This model is derived from the infrared luminosity functions of star forming galaxies, broken into two groups: `main sequence' galaxies and `starburst galaxies', combined with new spectral energy distributions from the {\it Herschel} observatory as well as source counts from a range of IR and submm wavelengths. The B{\'e}thermin model was scaled to $1.4\,$GHz assuming a non-evolving IR-radio correlation of \mbox{$q_{\rm TIR}\equiv \log\left({{L_{\rm IR}}\over{3.75\times10^{12}{\rm W}}}\times{{\rm W Hz^{-1}}\over{L_{1.4}}}\right)=2.64$} out to high redshift and a spectral index of $\alpha=0.8$. This model is plotted in Fig.~\ref{fig:infrared}, with our best fit results, as well as the Condon (1984) model and \citetalias{Condon12} power law. In contrast to the Condon (1984) model, the B{\'e}thermin model matches our results quite closely in the region $1 \leq S \, (\mu{\rm Jy}) \leq 50$, where the star-forming contribution is dominant. However, for $50 \leq S \, (\mu{\rm Jy}) \leq 1000$ the model drops below our best fits, as well as the Condon (1984) model, and is clearly under-predicting the observed counts. From figure 3 of \citet{Bethermin12b} this is the region where the main sequence contribution starts to decline and where the starburst contribution peaks. If the starburst contribution is increased by a factor of around 3 then the model in this region more closely approximates the other estimates.

\begin{figure}
\includegraphics[scale=0.375,natwidth=9in,natheight=9in]{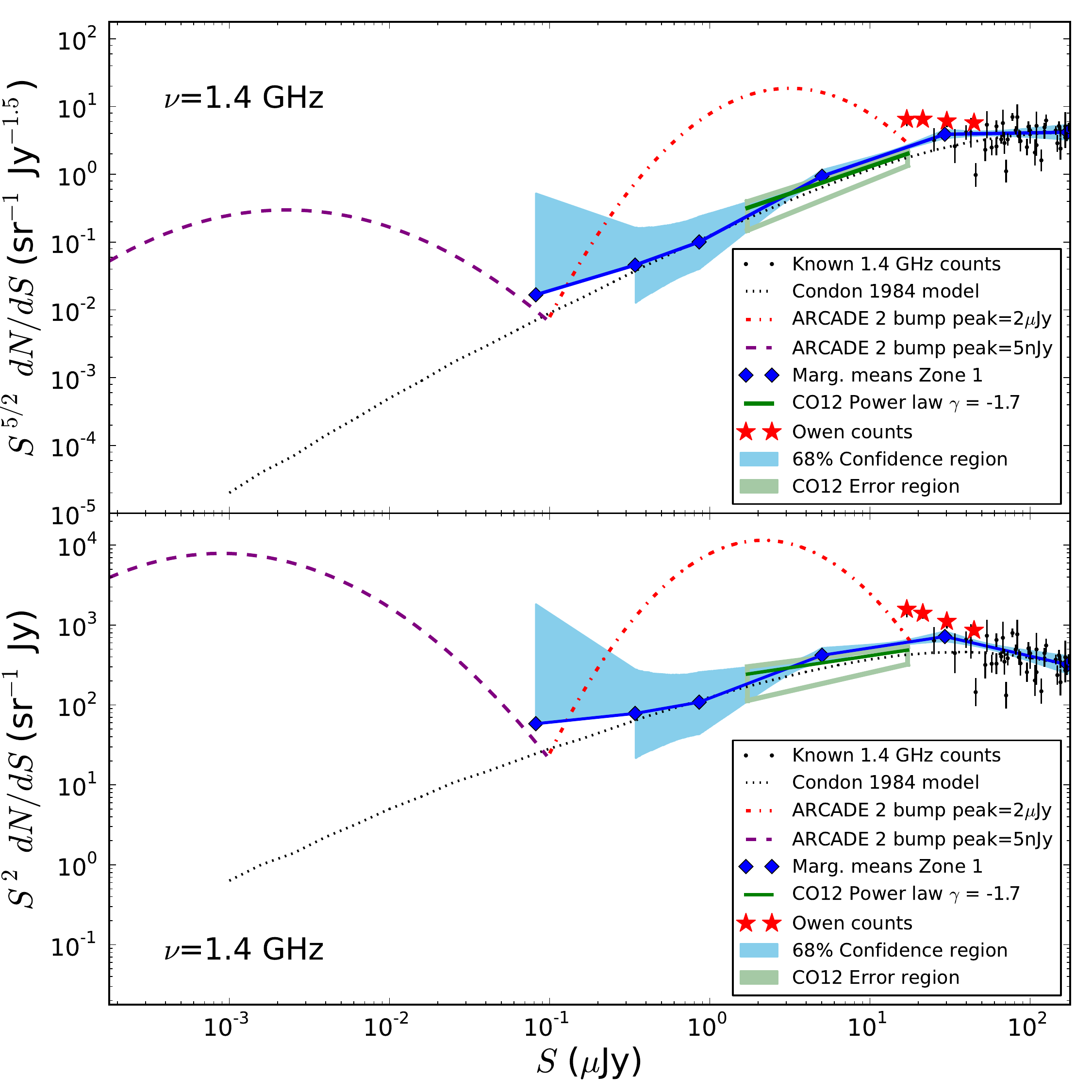}
\caption{Faint end of the $1.4 \,$GHz source count. The top panel is with the Euclidean normalisation, while the bottom panel is $S^2$ normalised. The bumps are two examples of counts that, when integrated, produce the extra background temperature necessary to match the ARCADE 2 emission, with the red dot-dashed line peaking at $2\,\mu$Jy and the purple dashed bump peaking at $5\,$nJy. The red star points show the source count of \citet{Owen08}. }
\label{fig:arcade}
\end{figure}

All of our model fits, even allowing liberal uncertainties, lie below the source-count values from \citet{Owen08}. These points are highlighted in red in Fig.~\ref{fig:arcade}; they seem to level off, or rise, toward fainter flux densities. We do not see any such indication for our results, all of our model fits declining in amplitude within this region and beyond. As discussed in \citetalias{Condon12} we believe this discrepancy to be mainly due to incorrect source size corrections in constructing the earlier source count estimates. Higher resolution VLA observations at $3\,$GHz would resolve the size issues definitively. 

As to the matter of the ARCADE 2 excess emission, it seems unlikely from these results that it could be coming from discrete sources. All of our model fits, both node-based and modified power law, as well as the single power-law from \citetalias{Condon12}, imply a background temperature at $3 \,$GHz of around $13\,$mK. Using the fit provided in \citet{Fixsen09} to scale the ARCADE 2 result from $3.2 \,$GHz to $3 \,$GHz yields a temperature of $62\,$mK, far outside the uncertainties in our results. 

There is no indication in our results of any new population of sources. Fig.~\ref{fig:arcade} shows two possible bumps (representing new possible populations) that would integrate up to the extra temperature necessary to account for the ARCADE 2 result. These were modelled as simple parabolas in the $\log_{10}[S]-\log_{10}[S^{5/2}dN/dS]$ plane, with fixed peak position. The bump peaking at around $2 \, \mu$Jy is clearly much higher in amplitude than any of our fits. Any kind of new population peaking above about $50 \,$nJy can be ruled out. Of course there is still the possibility that a new population could exist that is even fainter than our current limits, peaking somewhere below $50 \,$nJy. The fainter bump shown in Fig.~\ref{fig:arcade} is one such example. However, the source density required for such sources to contribute significantly to the background is extreme.

Between the faintest two nodes, $0.05\, \mu$Jy and $0.20\,\mu$Jy, and particularly near the faintest node, the count is not well constrained and the uncertainties do allow for a rise in the count. We therefore cannot rule out bumps with peaks fainter than $10\,$nJy. However, as the peak goes to fainter flux densities the bump needs to increase in width or height to produce the required background temperature. Note that any such population would far exceed the total number of known galaxies, as well as requiring a complete departure from the radio/far-IR correlation. 

\begin{table}
\centering
  \caption{Marginalised parameter means for the best fit (three-zone) node model and modified power law model scaled to $1.4\,$GHz using $\langle\alpha\rangle = -0.7$. The form for the modified power law is given in eq.~(\ref{eq:powerl})  }
  \begin{tabular}{cc}
\hline 
\hline
Node $1.4\,$GHz &Marginalised means\\
$\mu$Jy&$\log_{10}$[sr$^{-1}$ Jy$^{-1}$]\\
\hline
0.08& $15.55^{+1.20}$\\
0.34 &   $14.82^{+0.45}_{-0.43} $\\
0.86 &  $14.20^{+0.20}_{-0.20}$\\
5.02 &  $13.24^{+0.03}_{-0.03}$\\
29.3 &  $11.87^{+0.02}_{-0.02}$\\
171. &  $10.11^{+0.02}_{-0.02}$\\
963   & 8.31\\
21600  & 6.08\\
\hline
Parameter $1.4 \,$GHz &Marginalised means\\
$0.08\le S \,( \mu{\rm Jy}) \le 100.$&\\
\hline
$\alpha$ & $-4.7^{+1.2}_{-1.2}$\\
$\beta$& $-0.16^{+0.25}_{-0.25}$ \\
$\gamma$ &$0.016^{+0.016}_{-0.016}$ \\
$\log_{10}(\kappa)$ &$-4.68^{+1.2}_{-1.1}$\\
\hline
\end{tabular}
\label{tab:bestfits14}
\end{table}

\section{Conclusions}
Our VLA image \citepalias{Condon12} is the deepest currently available, with an instrumental noise of $1\, \mu$Jy beam$^{-1}$. To do justice to these data we have developed a novel and thorough \textit{P(D)} analysis that has revealed the structure of the $3$-GHz source count down to $0.1\, \mu$Jy.

The novel features are the following.
\begin{enumerate}
\item
We have modelled the source count by a series of nodes joined by short sections of power-law form \citep{Patanchon09}. In this way, there is no prescription, assumption or constraint 
on the form the count might follow. The parameters in our model then simply become the node values.
\item
We have used Markov chain Monte Carlo sampling throughout to provide unbiased determinations of the parameters and accurate estimates of parameter uncertainties. This demonstrates with clarity the dependence on flux density, how the inter-parameter dependencies increase with decreasing flux density, and the faintest limits to which \textit{P(D)} is sensitive.
\end{enumerate}

From the use of these novel techniques we have drawn the following conclusions.
\begin{enumerate}
\item
The MCMC approach shows that the uncertainties are dominated by sample variance rather than systematic effects, at least at the high end of the count. Hence a wider image \textit{at the same depth} would lead to an improved estimate of the source count.
\item
Our results are broadly consistent with the single power-law slope of $-1.7$ found by \citetalias{Condon12}, although differing slightly in detail. They show that the error estimate of \citetalias{Condon12} is somewhat generous. They also show with greater conviction the change to a shallower slope below $3\, \mu$Jy suggested by \citetalias{Condon12}.
\item
The consistency with previous estimates persists even when we take into account changes in the instrumental noise with frequency and position within the primary beam, different weightings of the wide-band bandpass data, and non-Gaussian features in the noise.
\item
We have shown that the method allows extraction of count information from these data to flux densities an order of magnitude below the limit traditionally set by noise plus confusion, and
far below the $5\sigma$ noise limit of around $5\, \mu$Jy set by direct source-counting. 
\item
Using a realistic large-scale simulation from \citet{Wilman08}, we have verified our approach and shown that it is unbiased. This simulation enabled us to quantify the effects of clustering and source sizes on the \textit{P(D)} distribution, both of which we found to be insignificant. While simulated \textit{P(D)} from a model sky is not new \citep[e.g.][]{Wall75}, never before has a comprehensive simulation been combined with a comprehensive count-fitting technique.
\item
Our source count estimates rule out any new populations that could be invoked to account for the ARCADE 2 excess temperature, down to a level of about $50\,$nJy. The count is closely represented by existing models of evolving luminosity functions, including the dominance of star-forming galaxies at the faintest flux densities observed; this suggests that we have a substantially robust accounting of the galaxies that contribute to the radio sky. 
\end{enumerate}

We have presented several different estimates of the count extracted directly from \textit{P(D)} analysis at $3\,$GHz, and the count extrapolated to $1.4\,$GHz. These involve different assumptions about the model shape, e.g. the number of nodes, and different choices of data, namely the various noise zones. Our most precise estimates come from the maximal use of the data (i.e. 3 noise zones) with 6 nodes, presented in the right column of Table~\ref{tab:params}. This choice allows sufficient freedom to fit the shape in each flux density region, without affecting the others. However, a simpler parameterization to use is provided by the modified power-law shape, which still fits the data reasonably well. The parameters for these fits are provided in Table~\ref{tab:modpl} at $3\,$GHz. The counts extrapolated to $1.4\,$GHz for both models are additionally provided in Table~\ref{tab:bestfits14}; this is particularly useful for comparison to this popular radio band where a large amount of data and models already exist.

\section*{Acknowledgments}
We acknowledge the support of the Natural Sciences and Engineering Research Council (NSERC) of Canada. We thank the staff of the VLA , which is operated by the National Radio Astronomy Observatory (NRAO). The National Radio Astronomy Observatory is a facility of the National Science Foundation operated under cooperative agreement by Associated Universities, Inc. We thank a diligent referee for helpful suggestions leading to improvement of the manuscript. The authors would also like to thank Gaelen Marsden for help implementing the \textit{P(D)} fitting procedure and for many useful discussions.

\bibliographystyle{mn2e}
\bibliography{mostupdatebib}

\begin{thebibliography}{}

\bibitem[\protect\citeauthoryear{{B{\'e}thermin}, {Daddi}, {Magdis}, {Sargent},
  {Hezaveh}, {Elbaz}, {Le Borgne}, {Mullaney}, {Pannella}, {Buat},
  {Charmandaris}, {Lagache} \& {Scott}}{{B{\'e}thermin}
  et~al.}{2012}]{Bethermin12b}
{B{\'e}thermin} M.,  {Daddi} E.,  {Magdis} G.,  {Sargent} M.~T.,  {Hezaveh} Y.,
   {Elbaz} D.,  {Le Borgne} D.,  {Mullaney} J.,  {Pannella} M.,  {Buat} V.,
  {Charmandaris} V.,  {Lagache} G.,    {Scott} D.,  2012, \apjl, 757, L23

\bibitem[\protect\citeauthoryear{{Blake}, {Mauch} \& {Sadler}}{{Blake}
  et~al.}{2004}]{Blake04}
{Blake} C.,  {Mauch} T.,    {Sadler} E.~M.,  2004, \mnras, 347, 787

\bibitem[\protect\citeauthoryear{{Blake} \& {Wall}}{{Blake} \&
  {Wall}}{2002a}]{Blake02a}
{Blake} C.,  {Wall} J.,  2002a, \mnras, 329, L37

\bibitem[\protect\citeauthoryear{{Blake} \& {Wall}}{{Blake} \&
  {Wall}}{2002b}]{Blake02b}
{Blake} C.,  {Wall} J.,  2002b, \mnras, 337, 993

\bibitem[\protect\citeauthoryear{{Chapin}, {Chapman} \& {Coppin}}{{Chapin}
  et~al.}{2011}]{Chapin11}
{Chapin} E.~L.,  {Chapman} S.~C.,    {Coppin} K.~E. e.~a.,  2011, \mnras, 411,
  505

\bibitem[\protect\citeauthoryear{{Condon}}{{Condon}}{1974}]{Condon74}
{Condon} J.~J.,  1974, \apj, 188, 279

\bibitem[\protect\citeauthoryear{{Condon}}{{Condon}}{1984}]{Condon84b}
{Condon} J.~J.,  1984, \apj, 287, 461

\bibitem[\protect\citeauthoryear{{Condon}, {Anderson} \& {Helou}}{{Condon}
  et~al.}{1991}]{Condon91}
{Condon} J.~J.,  {Anderson} M.~L.,    {Helou} G.,  1991, \apj, 376, 95

\bibitem[\protect\citeauthoryear{{Condon}, {Cotton}, {Fomalont}, {Kellermann},
  {Miller}, {Perley}, {Scott}, {Vernstrom} \& {Wall}}{{Condon}
  et~al.}{2012}]{Condon12}
{Condon} J.~J.,  {Cotton} W.~D.,  {Fomalont} E.~B.,  {Kellermann} K.~I.,
  {Miller} N.,  {Perley} R.~A.,  {Scott} D.,  {Vernstrom} T.,    {Wall} J.~V.,
  2012, \apj, 758, 23

\bibitem[\protect\citeauthoryear{{Condon}, {Cotton}, {Greisen}, {Yin},
  {Perley}, {Taylor} \& {Broderick}}{{Condon} et~al.}{1998}]{Condon98}
{Condon} J.~J.,  {Cotton} W.~D.,  {Greisen} E.~W.,  {Yin} Q.~F.,  {Perley}
  R.~A.,  {Taylor} G.~B.,    {Broderick} J.~J.,  1998, \aj, 115, 1693

\bibitem[\protect\citeauthoryear{{Cotton}}{{Cotton}}{2008}]{Cotton08}
{Cotton} W.~D.,  2008, \pasp, 120, 439

\bibitem[\protect\citeauthoryear{{de Jong}, {Klein}, {Wielebinski} \&
  {Wunderlich}}{{de Jong} et~al.}{1985}]{deJong85}
{de Jong} T.,  {Klein} U.,  {Wielebinski} R.,    {Wunderlich} E.,  1985, \aap,
  147, L6

\bibitem[\protect\citeauthoryear{{de Zotti}, {Massardi}, {Negrello} \&
  {Wall}}{{de Zotti} et~al.}{2010}]{dezotti09}
{de Zotti} G.,  {Massardi} M.,  {Negrello} M.,    {Wall} J.,  2010, \aapr, 18,
  1

\bibitem[\protect\citeauthoryear{{Fixsen}, {Kogut}, {Levin}, {Limon}, {Lubin},
  {Mirel}, {Seiffert}, {Singal}, {Wollack}, {Villela} \& {Wuensche}}{{Fixsen}
  et~al.}{2009}]{Fixsen09}
{Fixsen} D.~J.,  {Kogut} A.,  {Levin} S.,  {Limon} M.,  {Lubin} P.,  {Mirel}
  P.,  {Seiffert} M.,  {Singal} J.,  {Wollack} E.,  {Villela} T.,    {Wuensche}
  C.~A.,  2009, arxiv:0901.0555

\bibitem[\protect\citeauthoryear{{Friedmann} \& {Bouchet}}{{Friedmann} \&
  {Bouchet}}{2004}]{Friedmann04}
{Friedmann} Y.,  {Bouchet} F.,  2004, \mnras, 348, 737

\bibitem[\protect\citeauthoryear{Glenn, Conley, Bethermin, Altieri, Amblard,
  Arumugam, Aussel, Babbedge, Blain, Bock, Boselli, Buat, Castro-Rodriguez,
  Cava, Chanial, Clements, Conversi, Cooray, Dowell, Dwek, Eales \& ...}{Glenn
  et~al.}{2010}]{Glenn10}
Glenn J.,  Conley A.,  Bethermin M.,  Altieri B.,  Amblard A.,  Arumugam V.,
  Aussel H.,  Babbedge T.,  Blain A.,  Bock J.,  Boselli A.,  Buat V.,
  Castro-Rodriguez N.,  Cava A.,  Chanial P.,  Clements D.~L.,  Conversi L.,
  Cooray A.,  Dowell C.~D.,  Dwek E.,  Eales S.,    ... D.~E.,  2010, arXiv,
  astro-ph.CO

\bibitem[\protect\citeauthoryear{{Haarsma} \& {Partridge}}{{Haarsma} \&
  {Partridge}}{1998}]{Haarsma98}
{Haarsma} D.~B.,  {Partridge} R.~B.,  1998, \apjl, 503, L5+

\bibitem[\protect\citeauthoryear{{Ivison}, {Alexander}, {Biggs}, {Brandt},
  {Chapin}, {Coppin}, {Devlin}, {Dickinson}, {Dunlop}, {Dye}, {Eales},
  {Frayer}, {Halpern}, {Hughes}, {Ibar}, {Kov{\'a}cs} \& {Marsden}}{{Ivison}
  et~al.}{2010}]{Ivison10}
{Ivison} R.~J.,  {Alexander} D.~M.,  {Biggs} A.~D.,  {Brandt} W.~N.,  {Chapin}
  E.~L.,  {Coppin} K.~E.~K.,  {Devlin} M.~J.,  {Dickinson} M.,  {Dunlop} J.,
  {Dye} S.,  {Eales} S.~A.,  {Frayer} D.~T.,  {Halpern} M.,  {Hughes} D.~H.,
  {Ibar} E.,  {Kov{\'a}cs} A.,    {Marsden} G.~.,  2010, \mnras, 402, 245

\bibitem[\protect\citeauthoryear{{Lewis} \& {Bridle}}{{Lewis} \&
  {Bridle}}{2002}]{Lewis02}
{Lewis} A.,  {Bridle} S.,  2002, \prd, 66, 103511

\bibitem[\protect\citeauthoryear{{Maloney}, {Glenn}, {Aguirre}, {Golwala},
  {Laurent}, {Ade}, {Bock}, {Edgington}, {Goldin}, {Haig}, {Lange}, {Mauskopf},
  {Nguyen}, {Rossinot}, {Sayers} \& {Stover}}{{Maloney}
  et~al.}{2005}]{Maloney05}
{Maloney} P.~R.,  {Glenn} J.,  {Aguirre} J.~E.,  {Golwala} S.~R.,  {Laurent}
  G.~T.,  {Ade} P.~A.~R.,  {Bock} J.~J.,  {Edgington} S.~F.,  {Goldin} A.,
  {Haig} D.,  {Lange} A.~E.,  {Mauskopf} P.~D.,  {Nguyen} H.,  {Rossinot} P.,
  {Sayers} J.,    {Stover} P.,  2005, \apj, 635, 1044

\bibitem[\protect\citeauthoryear{{Mills}}{{Mills}}{1952}]{Mills52}
{Mills} B.~Y.,  1952, Australian Journal of Scientific Research A Physical
  Sciences, 5, 266

\bibitem[\protect\citeauthoryear{{Owen} \& {Morrison}}{{Owen} \&
  {Morrison}}{2008}]{Owen08}
{Owen} F.~N.,  {Morrison} G.~E.,  2008, \aj, 136, 1889

\bibitem[\protect\citeauthoryear{{Patanchon}, {Ade}, {Bock}, {Chapin},
  {Devlin}, {Dicker}, {Griffin}, {Gundersen}, {Halpern} \&
  {Hargrave}}{{Patanchon} et~al.}{2009}]{Patanchon09}
{Patanchon} G.,  {Ade} P.~A.~R.,  {Bock} J.~J.,  {Chapin} E.~L.,  {Devlin}
  M.~J.,  {Dicker} S.~R.,  {Griffin} M.,  {Gundersen} J.~O.,  {Halpern} M.,
  {Hargrave} P.,  2009, \apj, 707, 1750

\bibitem[\protect\citeauthoryear{{Ryle} \& {Scheuer}}{{Ryle} \&
  {Scheuer}}{1955}]{Ryle55a}
{Ryle} M.,  {Scheuer} P.~A.~G.,  1955, Royal Society of London Proceedings
  Series A, 230, 448

\bibitem[\protect\citeauthoryear{{Scheuer}}{{Scheuer}}{1957}]{Scheuer57}
{Scheuer} P.~A.~G.,  1957, Proceedings of the Cambridge Philosophical Society,
  53, 764

\bibitem[\protect\citeauthoryear{{Scheuer}}{{Scheuer}}{1974}]{Scheuer74}
{Scheuer} P.~A.~G.,  1974, \mnras, 166, 329

\bibitem[\protect\citeauthoryear{{Seiffert}, {Fixsen}, {Kogut}, {Levin},
  {Limon}, {Lubin}, {Mirel}, {Singal}, {Villela}, {Wollack} \&
  {Wuensche}}{{Seiffert} et~al.}{2009}]{Seiffert09}
{Seiffert} M.,  {Fixsen} D.~J.,  {Kogut} A.,  {Levin} S.~M.,  {Limon} M.,
  {Lubin} P.~M.,  {Mirel} P.,  {Singal} J.,  {Villela} T.,  {Wollack} E.,
  {Wuensche} C.~A.,  2009, arxiv:0901.0559

\bibitem[\protect\citeauthoryear{{Singal}, {Stawarz}, {Lawrence} \&
  {Petrosian}}{{Singal} et~al.}{2010}]{Singal10}
{Singal} J.,  {Stawarz} {\L}.,  {Lawrence} A.,    {Petrosian} V.,  2010,
  \mnras, pp 1458--+

\bibitem[\protect\citeauthoryear{Takeuchi \& Ishii}{Takeuchi \&
  Ishii}{2004}]{Takeuchi04}
Takeuchi T.~T.,  Ishii T.~T.,  2004, ApJ, 604, 40

\bibitem[\protect\citeauthoryear{{Takeuchi}, {Kawabe}, {Kohno}, {Nakanishi},
  {Ishii}, {Hirashita} \& {Yoshikawa}}{{Takeuchi} et~al.}{2001}]{Takeuchi01}
{Takeuchi} T.~T.,  {Kawabe} R.,  {Kohno} K.,  {Nakanishi} K.,  {Ishii} T.~T.,
  {Hirashita} H.,    {Yoshikawa} K.,  2001, \pasp, 113, 586

\bibitem[\protect\citeauthoryear{{Vernstrom}, {Scott} \& {Wall}}{{Vernstrom}
  et~al.}{2011}]{Vernstrom11}
{Vernstrom} T.,  {Scott} D.,    {Wall} J.~V.,  2011, \mnras, 415, 3641

\bibitem[\protect\citeauthoryear{{Wall} \& {Cooke}}{{Wall} \&
  {Cooke}}{1975}]{Wall75}
{Wall} J.~V.,  {Cooke} D.~J.,  1975, \mnras, 171, 9

\bibitem[\protect\citeauthoryear{Wilman, Miller, Jarvis, Mauch, Levrier,
  Abdalla, Rawlings, Kl{\"o}ckner, Obreschkow, Olteanu \& Young}{Wilman
  et~al.}{2008}]{Wilman08}
Wilman R.~J.,  Miller L.,  Jarvis M.~J.,  Mauch T.,  Levrier F.,  Abdalla
  F.~B.,  Rawlings S.,  Kl{\"o}ckner H.-R.,  Obreschkow D.,  Olteanu D.,
  Young S.,  2008, Monthly Notices of the Royal Astronomical Society, 388, 1335

\end{thebibliography}

\bsp

\label{lastpage}
\end{document}